# Strongly correlated electronic superconductivity in the noncentrosymmetric Re-Os-based high/medium-entropy alloys

Rui Chen[1], Longfu Li[1], Lingyong Zeng[1,2], Kuan Li[1], Peifeng Yu[1], Kangwang Wang[1], Zaichen Xiang[1], Shuangyue Wang[1], Jingjun Qin[1], Wanyi Zhang[1], Yucheng Li[1], Tian Shang[3], Huixia Luo[1, 4,5,6*]

[1]*School of Materials Science and Engineering, Sun Yat-sen University, No. 135, Xingang Xi Road, Guangzhou, 510275, P. R. China*

[2]*Device Physics of Complex Materials, Zernike Institute for Advanced Materials, University of Groningen, Nijenborgh 4, 9747 AG Groningen, The Netherlands*

[3]*Key Laboratory of Polar Materials and Devices (MOE), School of Physics and Electronic Science, East China Normal University, Shanghai 200241, China*

[4]*State Key Laboratory of Optoelectronic Materials and Technologies, Sun Yat-sen University, No. 135, Xingang Xi Road, Guangzhou, 510275, P. R. China*

[5]*Key Lab of Polymer Composite & Functional Materials, Sun Yat-sen University, No. 135, Xingang Xi Road, Guangzhou, 510275, P. R. China*

[6]*Guangdong Provincial Key Laboratory of Magnetoelectric Physics and Devices, Sun Yat-sen University, No. 135, Xingang Xi Road, Guangzhou, 510275, P. R. China*

[*]*Corresponding author. Email: luohx7@mail.sysu.edu.cn (H. Luo)*

**Abstract**

The class of unconventional superconductors, particularly noncentrosymmetric superconductors, has been highly considered as potential materials for understanding the complex properties of quantum materials. Here, five previously unreported $Re_{3.5}Os_{3.5}Ta_{0.5}Hf_{0.5}Nb_3$, $Re_3Os_3Ta_{0.5}Hf_{0.5}Nb_3$, $Re_{3.5}Os_{3.5}Mo_{0.5}Hf_{0.5}Nb_3$, $Re_{3.5}Os_{3.5}Mo_{0.5}W_{0.5}Nb_3$, and $Re_3Os_3Mo_{0.5}Hf_{0.5}Nb_3$ Re-Os-based high/medium-entropy alloys (MEAs-HEAs) with valence electron count ranging from 6.45 to 6.81 were synthesized and investigated using x-ray diffraction, transport, magnetization, and specific heat measurements. Our analyses confirm that all five compounds crystallize in a noncentrosymmetric $\alpha$-Mn-type structure and exhibit type-II superconductivity with $T_c$ values from 4.20 K to 5.11 K, respectively. Unexpectedly, despite being immersed in an acidic environment for one month, the structures and superconducting properties of HEAs remain stable. Our findings indicate that the $T_c$ increases with an increasing valence electron count in MEAs-HEAs. Furthermore, these noncentrosymmetric $\alpha$-Mn-type HEA superconductors have large Kadowaki-Woods ratios (KWR), implying the presence of strong electronic correlations.

## 1. Introduction

With the rapid development of technology, there is a great demand for developing advanced materials with enhanced performance [1-5]. For example, traditional metal alloys have gradually become a bottleneck, especially under certain extreme loads or environments, prompting researchers to ceaselessly explore new materials. Medium/ high-entropy alloys (MEAs-HEAs) offer a novel path for the design of advanced materials because of the combination of various remarkable properties, such as high strength, high toughness, and superior ductility [6-8]. In addition, the chemical complexity of MEAs-HEAs endows them with extensive tunability in both structure and properties, thereby providing a fertile platform to investigate the interplay between structure and properties in multicomponent alloy systems [2,9].

Since the first discovery of superconductivity in HEAs in 2014 [10], many MEAs-HEAs have been identified as superconductors, with the majority exhibiting simple centrosymmetric structural types, such as body-centered-cubic (BCC) type [11-13], hexagonal-closed packed (HCP) type [14,15], and face-centered-cubic type (FCC) [16,17]. However, the exploration of noncentrosymmetric (NC) MEA-HEA superconductors is mostly limited to the $\alpha$-Mn [18,19] and $\beta$-Mn [20-24] type structure alloys.

On the other hand, a large number of noncentrosymmetric superconductors (NCS) have been previously confirmed to exhibit various unique properties, such as nodes in the superconducting gap [20,25,26], upper critical field exceeding Pauli paramagnetic limit [27,28], and time-reversal symmetry breaking (TRSB) [29,30]. The first reported NC $\alpha$-Mn-type HEA superconductors are $(HfTaWIr)_{1-x}Re_x$, $(ZrNb)_{1-x}(MoReRu)_x$, and $(HfTaWPt)_{1-x}Re_x$, which display strong composition dependencies of superconducting transition temperatures ($T_c$s) [31]. Furthermore, the upper critical field of the NC $\alpha$-Mn type $Re_{0.35}Os_{0.35}Mo_{0.10}W_{0.10}Zr_{0.10}$ is found to exceed the Pauli limiting field, implying that unconventional superconductivity may exist [32]. Therefore, it will be interesting to explore new MEAs-HEAs superconductors with an NC structure could be interesting.

Understanding the role of crystal structure, element composition, local microstructure, and spin-orbit coupling (SOC) strength is essential for studying the superconducting properties and pairing mechanism of the Re-Os system [33]. At present, the research of Re-Os-based superconductors mainly focuses on binary alloys containing transition metals ($T$), in which the composition changes of Re-Os and $T$ site lead to the crystallization in the crystal structure of centrosymmetry or non-centrosymmetry. In binary $Re_{1-x}Mo_x$ alloy, by changing the Re/Mo ratio, there are four different solid phases: hcp-Mg, $\alpha$-Mn, $\beta$-CrFe, and bcc-W, of which the second is NC and the rest are centrosymmetric [34]. The ratio of Re to $T$ may be the factor leading to the fracture of the time-reversal symmetry. In order to fully understand the exact pairing mechanism, it is necessary to further study and micro-measure the new Re-Os-based superconductors. The Re-Os-based MEAs-HEAs offer an excellent opportunity for this purpose due to their mixed 4$d$/5$d$ site. Compared to the binary analog, Re-Os-based MEAs-HEAs introduces additional disorder due to the variation in the atomic size of Ta, Nb, Mo, and W.

In this study, five previously unreported Re-Os-based NC $\alpha$-Mn-type $Re_{3.5}Os_{3.5}Ta_{0.5}Hf_{0.5}Nb_3$, $Re_3Os_3Ta_{0.5}Hf_{0.5}Nb_3$, $Re_{3.5}Os_{3.5}Mo_{0.5}Hf_{0.5}Nb_3$, $Re_{3.5}Os_{3.5}Mo_{0.5}W_{0.5}Nb_3$, and $Re_3Os_3Mo_{0.5}Hf_{0.5}Nb_3$ MEA-HEA compositions are designed by adjusting the valence electron count (VEC) from 6.45 to 6.81. Our analyses confirm that these MEAs-HEAs exhibit type-II superconductivity with $T_c$ ranging from 4.20 K to 5.11 K. Moreover, to investigate superconducting MEAs-HEAs to acidic environmental corrosion [35,36], they are immersed in an HCl solution for one month. Furthermore, by comprehensively investigating these superconductors, we aim to elucidate the intricacies of $T_c$ and VEC and the strength of electron-electron correlations in the NC $\alpha$-Mn-type HEA superconductors.

## 2. Experimental details

$Re_{3.5}Os_{3.5}Ta_{0.5}Hf_{0.5}Nb_3$, $Re_3Os_3Ta_{0.5}Hf_{0.5}Nb_3$, $Re_{3.5}Os_{3.5}Mo_{0.5}Hf_{0.5}Nb_3$, $Re_{3.5}Os_{3.5}Mo_{0.5}W_{0.5}Nb_3$, and $Re_3Os_3Mo_{0.5}Hf_{0.5}Nb_3$ were synthesized by arc melting of high-purity Re, Os, Mo, Hf, Nb, and Ta in an argon atmosphere, with multiple careful melting cycles to ensure uniformity**.** To determine the phase purity and crystal structure of these samples, X-ray diffraction (XRD) measurements were performed by employing the MiniFlex of Rigaku equipped with Cu Kα radiation, and the Rietveld method was used to determine the crystal structure using Fullprof Suite software. The elemental ratios were examined by scanning electron microscope (SEM) combined with energy-dispersive x-ray spectroscopy (EDS). Transport, magnetization, and specific heat measurements were measured in a Quantum Design physical property measurement system PPMS-14T.

## 3. Results and discussion

### 3.1. structure characterization

Rietveld refined the powder XRD patterns using the FullProf Suite software, which are shown in **Figure 1a**. XRD results confirm that five alloys have a single phase with $\alpha$-Mn structure (space group, $I$-43m). **Table 1** summarizes the lattice parameters for the five alloys. The refined lattice parameters are $a$ = 9.58 Å for $Re_{3.5}Os_{3.5}Ta_{0.5}Hf_{0.5}Nb_3$, $a$ = 9.82 Å for $Re_3Os_3Ta_{0.5}Hf_{0.5}Nb_3$, $a$ = 9.56 Å for $Re_{3.5}Os_{3.5}Mo_{0.5}Hf_{0.5}Nb_3$, $a$ = 10.42 Å for $Re_3Os_3Mo_{0.5}Hf_{0.5}Nb_3$, and $a$ = 9.76 Å for $Re_{3.5}Os_{3.5}Mo_{0.5}W_{0.5}Nb_3$. **Table S1** summarizes tabulated reliability factors ($R_{wp}$, $R_p$, $\chi^2$). **Figure 1b** shows the powder XRD of pristine $Re_{3.5}Os_{3.5}Ta_{0.5}Hf_{0.5}Nb_3$, $Re_3Os_3Ta_{0.5}Hf_{0.5}Nb_3$, $Re_{3.5}Os_{3.5}Mo_{0.5}Hf_{0.5}Nb_3$, $Re_{3.5}Os_{3.5}Mo_{0.5}W_{0.5}Nb_3$, and $Re_3Os_3Mo_{0.5}Hf_{0.5}Nb_3$ MEAs-HEAs and these five MEAs-HEAs immersed in HCl solution for one month, respectively. The positions of the diffraction peaks of $Re_3Os_3Ta_{0.5}Hf_{0.5}Nb_3$ and $Re_3Os_3Mo_{0.5}Hf_{0.5}Nb_3$ have no significant changes compared to other MEAs-HEAs, even after immersion for one month, the structural integrity of these alloys remains highly stable in a hydrochloric acid (HCl) environment.

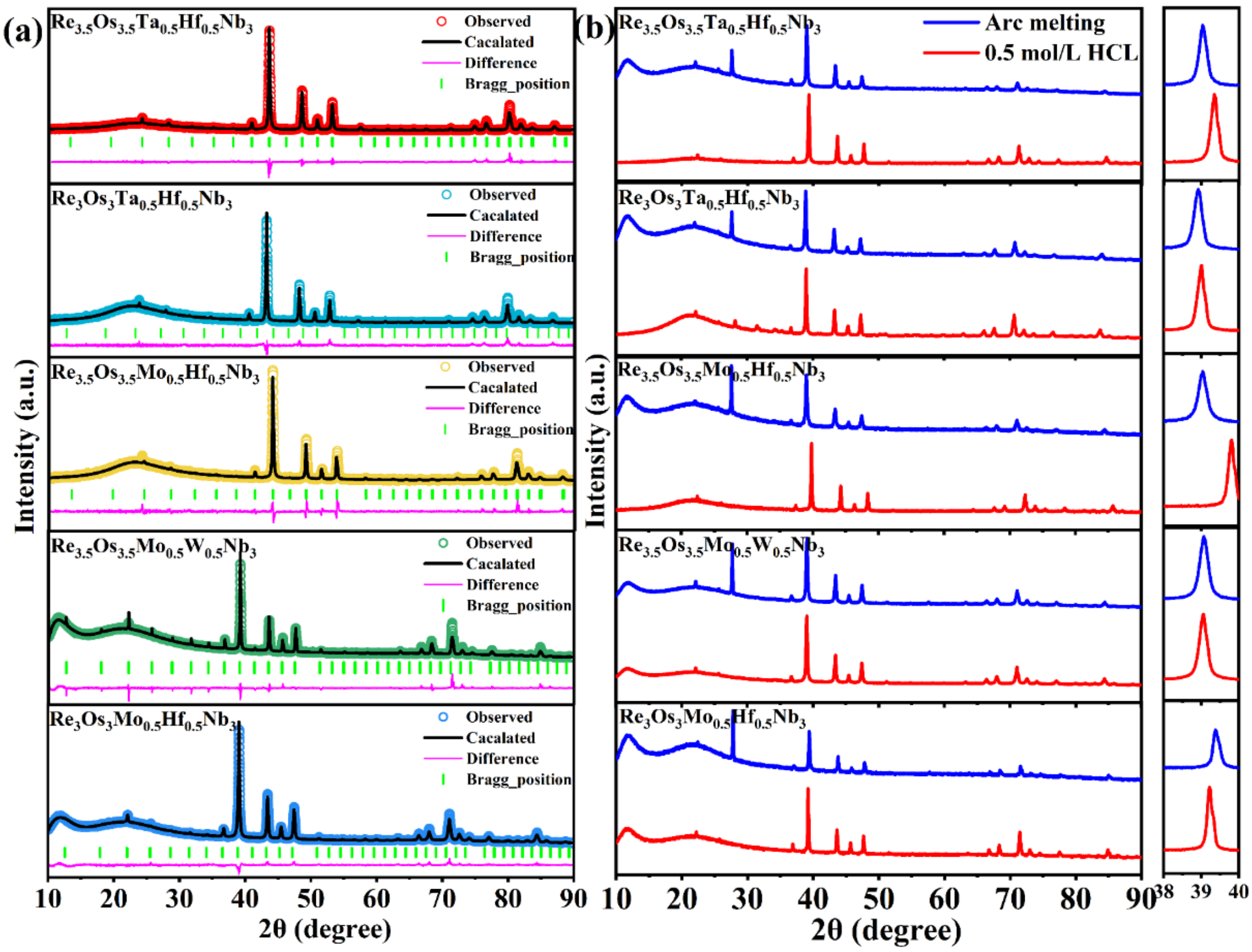

**Figure 1.** (a) The pristine powder XRD patterns for these five $Re_{3.5}Os_{3.5}Ta_{0.5}Hf_{0.5}Nb_3$, $Re_3Os_3Ta_{0.5}Hf_{0.5}Nb_3$, $Re_{3.5}Os_{3.5}Mo_{0.5}Hf_{0.5}Nb_3$, $Re_3Os_3Mo_{0.5}Hf_{0.5}Nb_3$, and $Re_{3.5}Os_{3.5}Mo_{0.5}W_{0.5}Nb_3$ samples along with Bragg positions. (b) The powder XRD of pristine MEAs-HEAs and MEAs-HEAs immersed in 0.5 mol/L HCl for one month, respectively.

**Table 1.** Starting atomic composition, theoretical VEC, structure parameters, chemical composition, and determined VEC of the five alloys.

| **Starting composition** | **Theoretical VEC** | ***a*** **(Å)** | **Determined composition** | **Determined VEC** |
|---|---|---|---|---|
| $Re_{3.5}Os_{3.5}Ta_{0.5}Hf_{0.5}Nb_3$ | 6.54 | 9.58 | $Re_{3.503}Os_{3.503}Ta_{0.492}Hf_{0.230}Nb_{2.272}$ | 6.65 |
| $Re_3Os_3Ta_{0.5}Hf_{0.5}Nb_3$ | 6.45 | 9.82 | $Re_{3.222}Os_{3.250}Ta_{0.537}Hf_{0.362}Nb_{2.629}$ | 6.58 |
| $Re_{3.5}Os_{3.5}Mo_{0.5}Hf_{0.5}Nb_3$ | 6.59 | 9.56 | $Re_{3.414}Os_{3.467}Mo_{0.387}Hf_{0.357}Nb_{2.375}$ | 6.72 |
| $Re_3Os_3Mo_{0.5}Hf_{0.5}Nb_3$ | 6.5 | 10.42 | $Re_{3.281}Os_{3.246}Mo_{0.52}Hf_{0.259}Nb_{2.694}$ | 6.70 |
| $Re_{3.5}Os_{3.5}Mo_{0.5}W_{0.5}Nb_3$ | 6.68 | 9.76 | $Re_{3.459}Os_{3.379}Mo_{0.5}W_{0.478}Nb_{2.158}$ | 6.79 |

**Figure S1-10** shows the SEM images of the five superconductors, which confirm that all samples exhibit no noticeable secondary phase and that the five elements have homogeneous distribution. The actual chemical compositions determined using EDS are all very close to the nominal compositions (**Table 1**). Furthermore, the EDS elemental mappings show all elements distributed evenly, and no clusters are seen. The actual ratios of the elements are in close agreement with the nominal ratios.

To uncover the surface corrosion of superconducting $Re_{3.5}Os_{3.5}Ta_{0.5}Hf_{0.5}Nb_3$, $Re_3Os_3Ta_{0.5}Hf_{0.5}Nb_3$, $Re_{3.5}Os_{3.5}Mo_{0.5}Hf_{0.5}Nb_3$, $Re_{3.5}Os_{3.5}Mo_{0.5}W_{0.5}Nb_3$, and $Re_3Os_3Mo_{0.5}Hf_{0.5}Nb_3$ MEAs-HEAs in the acidic environment, HEAs were characterized using SEM. For these MEAs-HEAs in HCl solution, there is almost a single α-Mn phase even after immersion in HCl for one month. All the elements are distributed uniformly. The ratio of elements in pristine MEAs-HEAs is consistent with that of MEAs-HEAs corroded for one month. Therefore, acidic corrosion of MEAs-HEAs will not result in a significant oxidation, while maintaining high quality within the MEAs-HEAs.

### 3.2. Resistivity

**Figures 2** depict the superconducting transitions reflected by the resistivity data, which exhibit $T_c$ at 4.52 K, 4.20 K, 4.94 K, 4.77 K, and 5.11 K for the pristine $Re_{3.5}Os_{3.5}Ta_{0.5}Hf_{0.5}Nb_3$, $Re_3Os_3Ta_{0.5}Hf_{0.5}Nb_3$, $Re_{3.5}Os_{3.5}Mo_{0.5}Hf_{0.5}Nb_3$, $Re_3Os_3Mo_{0.5}Hf_{0.5}Nb_3$, and $Re_{3.5}Os_{3.5}Mo_{0.5}W_{0.5}Nb_3$, respectively. The calculated residual resistivity ratios (RRRs) are 1.02, 1.05, 1.04, 1.04, and 1.01 for the pristine $Re_{3.5}Os_{3.5}Ta_{0.5}Hf_{0.5}Nb_3$, $Re_3Os_3Ta_{0.5}Hf_{0.5}Nb_3$, $Re_{3.5}Os_{3.5}Mo_{0.5}Hf_{0.5}Nb_3$, $Re_3Os_3Mo_{0.5}Hf_{0.5}Nb_3$, and $Re_{3.5}Os_{3.5}Mo_{0.5}W_{0.5}Nb_3$ samples. The RRRs are near 1, which is similar to other *α*-Mn compounds $Nb_{0.18}Re_{0.82}$ (RRR ≈ 1.3), $Re_{0.35}Os_{0.35}Mo_{0.10}W_{0.10}Zr_{0.10}$ (RRR ≈ 1.3)[32], $Re_6Zr$ (RRR ≈ 1.10)[37].

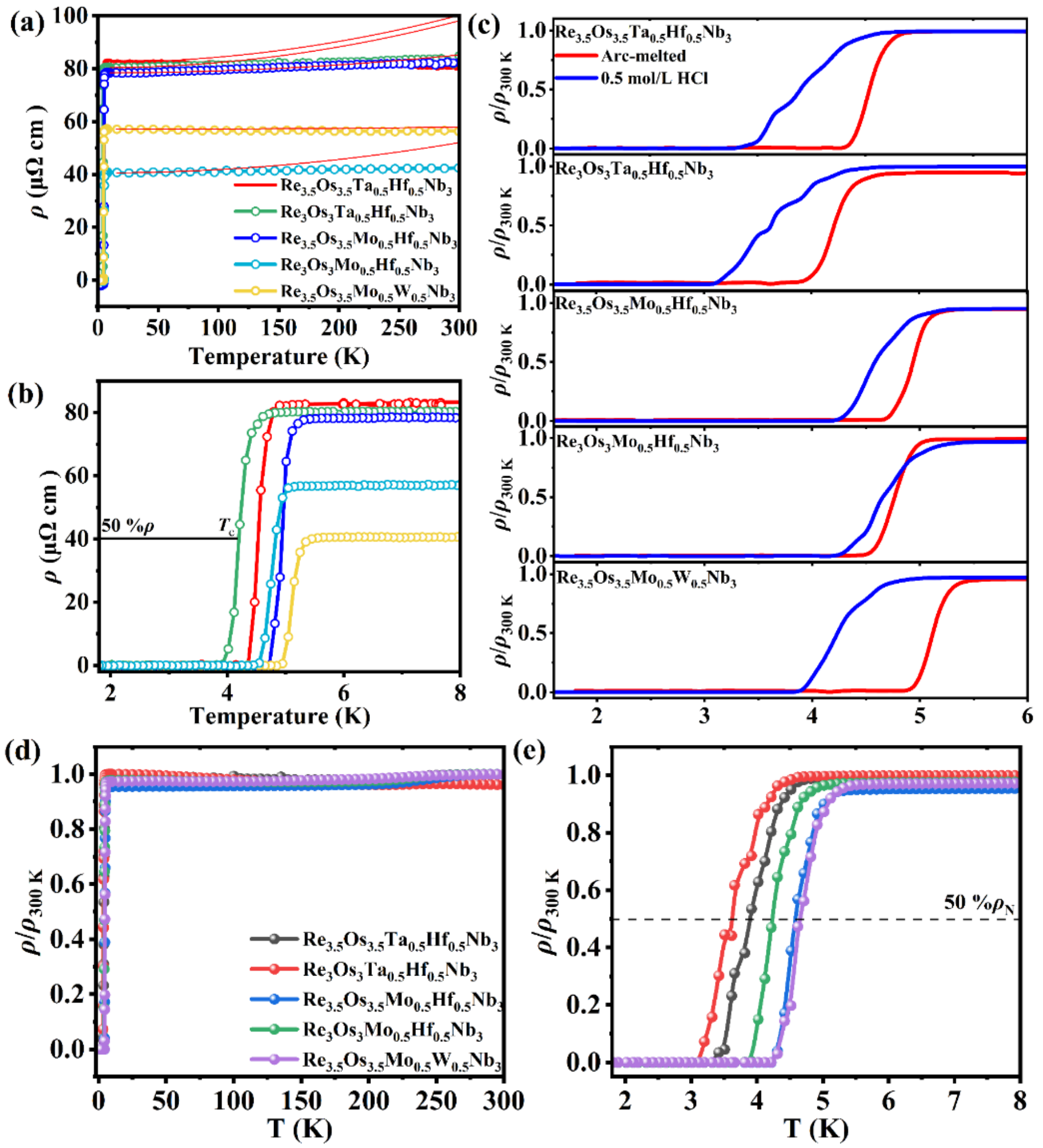

**Figure 2** (a) $\rho$(T) for pristine alloys, the red solid line represents a power law fitting. (b) The sharp 4.52 K, 4.20 K, 4.94 K, 4.77 K, and 5.11 K for our fresh $Re_{3.5}Os_{3.5}Ta_{0.5}Hf_{0.5}Nb_3$, $Re_3Os_3Ta_{0.5}Hf_{0.5}Nb_3$, $Re_{3.5}Os_{3.5}Mo_{0.5}Hf_{0.5}Nb_3$, $Re_3Os_3Mo_{0.5}Hf_{0.5}Nb_3$, and $Re_{3.5}Os_{3.5}Mo_{0.5}W_{0.5}Nb_3$ MEAs-HEAs. (c) Temperature-dependent sheet resistances of our pristine MEA-HEA superconductors and the spent MEAs-HEAs after immersing in the HCl solution for one month, respectively. (d) Temperature-dependent resistances of our five MEA-HEA after immersing for one month, respectively. (e) The enlarged view around $T_c$.

The normal-state resistivity $\rho$(T) exhibits almost temperature-independent behavior. We fit the low-temperature resistivity data from 10 K to 50 K by using a power law,

$$\rho(T) = \rho_0 + AT^n$$

This is shown in **Figure 2a** by the solid red line. The $\rho_0$ is residual resistivity, $A$ is the electronic contribution of electron-electron scattering to the resistivity at low temperatures in the Fermi-liquid picture. The best fit to the data is achieved for $n$ = 2, which yields $\rho_0$ = 823.99, 803.26, 783.95, 406.90, 570.68 $\mu\Omega$·cm and $A$ = 1.97 × $10^{-3}$, 1.97 × $10^{-3}$, 7.57 × $10^{-4}$, 1.27 × $10^{-3}$, 2.35 × $10^{-4}$ $\mu\Omega$ cm $K^{-2}$ for $Re_{3.5}Os_{3.5}Ta_{0.5}Hf_{0.5}Nb_3$, $Re_3Os_3Ta_{0.5}Hf_{0.5}Nb_3$, $Re_{3.5}Os_{3.5}Mo_{0.5}Hf_{0.5}Nb_3$, $Re_3Os_3Mo_{0.5}Hf_{0.5}Nb_3$, and $Re_{3.5}Os_{3.5}Mo_{0.5}W_{0.5}Nb_3$, respectively.

To investigate the stability of superconducting properties in the acid environment, $Re_{3.5}Os_{3.5}Ta_{0.5}Hf_{0.5}Nb_3$, $Re_3Os_3Ta_{0.5}Hf_{0.5}Nb_3$, $Re_{3.5}Os_{3.5}Mo_{0.5}Hf_{0.5}Nb_3$, $Re_3Os_3Mo_{0.5}Hf_{0.5}Nb_3$, and $Re_{3.5}Os_{3.5}Mo_{0.5}W_{0.5}Nb_3$ MEAs-HEAs were immersed in 0.5 mol/L HCl for one month. **Figure 2c** shows the temperature-dependence of the normalized $\rho/\rho_{300\ K}$ of Re-Os based MEAs-HEAs after immersing into 0.5 mol/L HCl for one month, together with the result of pristine MEAs-HEAs for comparison. Now the normalized resistivity value ($\rho_N$) decreases by 50 % from the normal state as the criterion of the $T_c$. It is noteworthy that compared with the original MEA-HEAs, the $T_c$ of these MEA-HEAs only slightly decreased after soaking in HCl solution for one month, indicating that superconducting MEA-HEAs have high resistance to acid corrosion. Remarkably, among these MEAs-HEAs, $Re_{3.5}Os_{3.5}Mo_{0.5}Hf_{0.5}Nb_3$ and $Re_3Os_3Mo_{0.5}Hf_{0.5}Nb_3$ retain excellent superconducting properties even after one month of immersion in the acidic environment (see **Figure 2d,e**). However, there are several shoulders in the resistance curves near critical points. Samples corroded by acidic solutions may exhibit phase separation due to differences in local composition or defect density, resulting in multiple shoulders on the resistance curve.

### 3.3 Magnetization

The zero-field-cooled magnetization (ZFC) measurements were performed at a magnetic field of 30 Oe, with the results shown in **Figure 3**. As depicted in **Figure 3**, the bulk superconductivity of the five MEA-HEAs is confirmed. The magnetic susceptibility is calculated by the formula: $\chi_v = \frac{M_v}{H}$, $H$ and $M_v$ represent the applied magnetic field and the volume magnetization [38]. Five samples exhibit a very large diamagnetic signal ($4\pi\chi_v \sim 1$) and the obtained $T_c$ = 3.91 K for $Re_{3.5}Os_{3.5}Ta_{0.5}Hf_{0.5}Nb_3$, $T_c$ = 3.42 K for $Re_3Os_3Ta_{0.5}Hf_{0.5}Nb_3$, $T_c$ = 4.39 K for $Re_{3.5}Os_{3.5}Mo_{0.5}Hf_{0.5}Nb_3$, $T_c$ = 4.14 K for $Re_3Os_3Mo_{0.5}Hf_{0.5}Nb_3$, and $T_c$ = 4.45 K for $Re_{3.5}Os_{3.5}Mo_{0.5}W_{0.5}Nb_3$, respectively. The magnetization curves are additionally adjusted by integrating the demagnetization factor (N) derived from the 1.8 K isothermal $M$(H) data. The shielding volume fractions approached 100 % at 1.8 K, which is indicative of the bulk superconductivity of five alloys[39]. To directly investigate the effect of HCl solution on superconductivity, we compared the ZFC magnetization of the pristine MEA-HEAs with that of MEA-HEAs immersed in a 0.5mol/L HCl solution for one month. As shown in **Figure 3**, among these MEAs-HEAs, the $T_c$ is almost unchanged, while the superconducting volume fraction is slightly reduced.

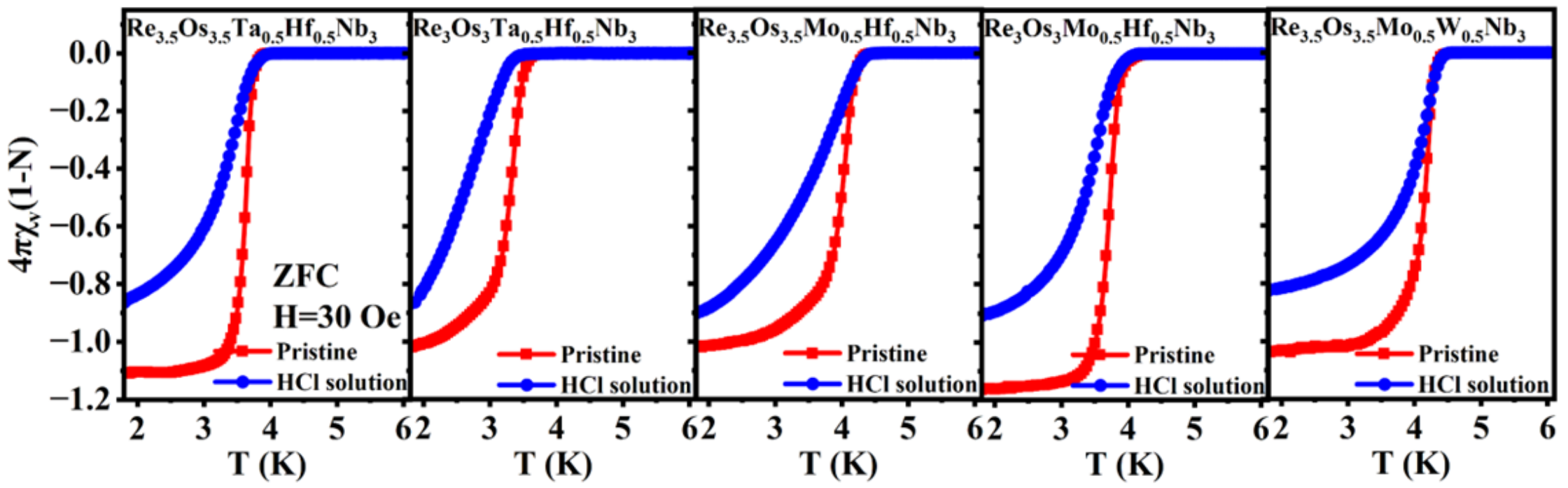

**Figure 3** ZFC magnetization at 30 Oe for pristine MEA-HEAs and after immersing in the 0.5 mol/L HCl solution for one month.

**Figure 4a-e** shows the $M$-$M_{fit}$ curves at different temperatures; the magnetic isotherms measured over a temperature range of 1.8 - 6 K are emphasized in the inset. For a perfect response to the field, $M_{fit} = m + n$H is determined by linear fitting of the magnetization data in the low-field region at 1.8 K[40]. The magnetization varies linearly when the magnetic field reaches a certain value, and the point at which it begins to deviate from the linear or Meissner line is regarded as the lower critical magnetic

field $\mu_0 H_{c1}^*$ at a particular temperature. The $\mu_0 H_{c1}^*$ values at 0 K are estimated by fitting the data points using the expression $\mu_0 H_{c1}^*(T) = \mu_0 H_{c1}^*(0)[1-(T/T_C)^2]$. As shown in **Figure 4(f)**, $\mu_0 H_{c1}^*$ at 0 K is extrapolated to be 35 mT, 53 mT, 53 mT, 70 mT, and 100 mT for $Re_{3.5}Os_{3.5}Ta_{0.5}Hf_{0.5}Nb_3$, $Re_3Os_3Ta_{0.5}Hf_{0.5}Nb_3$, $Re_{3.5}Os_{3.5}Mo_{0.5}Hf_{0.5}Nb_3$, $Re_3Os_3Mo_{0.5}Hf_{0.5}Nb_3$, and $Re_{3.5}Os_{3.5}Mo_{0.5}W_{0.5}Nb_3$, respectively.

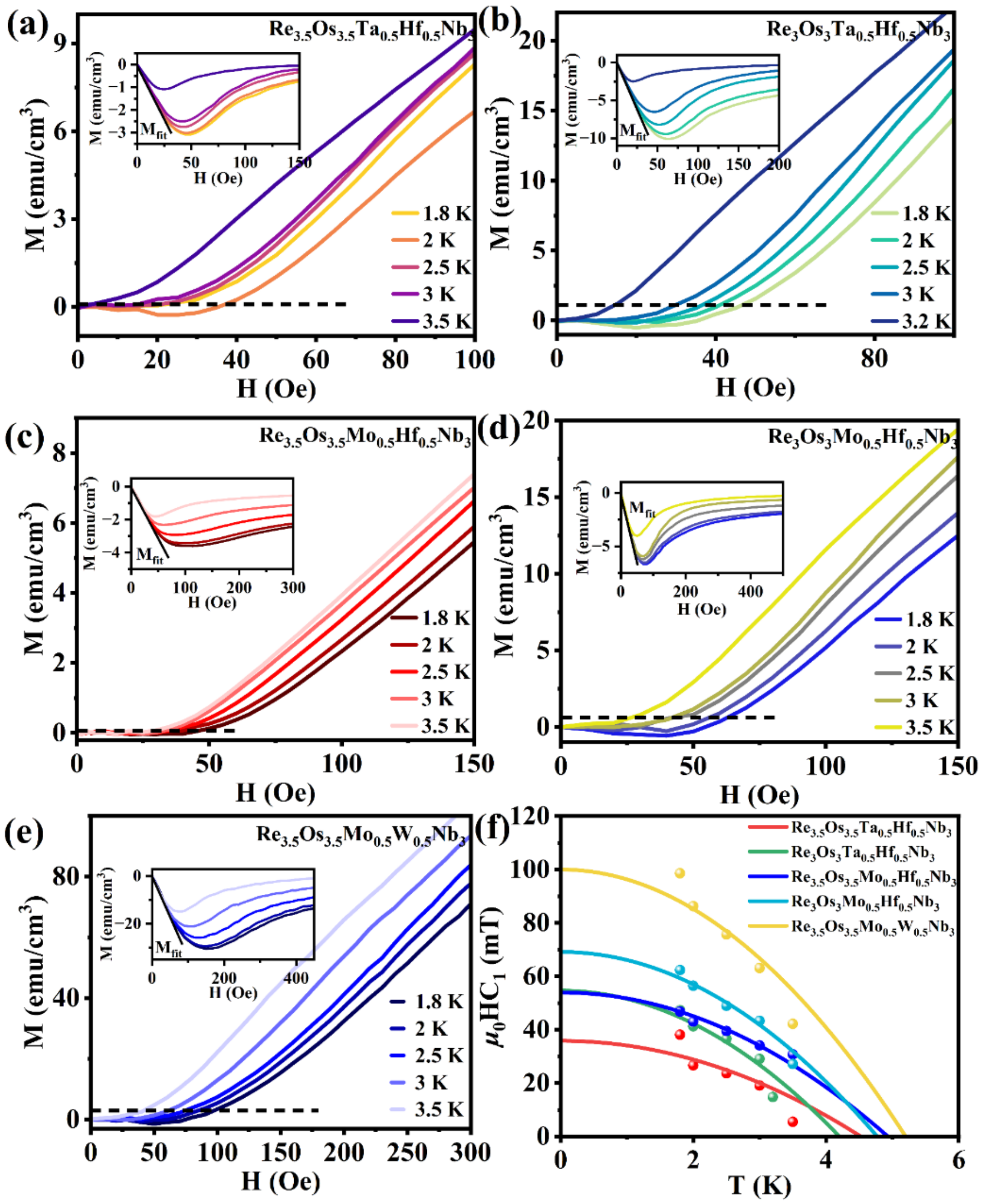

**Figure 4** (a-e) The *M*-$M_{\mathrm{fit}}$ curves at different temperatures, the insets display field-dependent magnetization. (f) The lower critical field versus.

Temperature-dependent electrical resistivity, as shown in **Figure 5**, was further investigated under various magnetic fields to estimate the upper critical field $\mu_0 H_{c2}$. When the applied external magnetic field increases, $T_c$ shifts to a lower temperature. $\mu_0 H_{c2}$ can be described using the Ginzburg-Landau (GL) equation $\mu_0 H_{c2}(T) = \mu_0 H_{c2}(0)\frac{1-(T/T_C)^2}{1+(T/T_C)^2}$. As the solid lines in **Figure 5f**, the estimated $\mu_0 H_{c2}{}^{GL}(0)$ values are 5.99 T, 7.71 T, 6.91 T, 6.29 T, and 6.69 T for $Re_{3.5}Os_{3.5}Ta_{0.5}Hf_{0.5}Nb_3$, $Re_3Os_3Ta_{0.5}Hf_{0.5}Nb_3$, $Re_{3.5}Os_{3.5}Mo_{0.5}Hf_{0.5}Nb_3$, $Re_3Os_3Mo_{0.5}Hf_{0.5}Nb_3$, and $Re_{3.5}Os_{3.5}Mo_{0.5}W_{0.5}Nb_3$, respectively. Remarkably, the zero-temperature upper critical fields of $Re_3Os_3Ta_{0.5}Hf_{0.5}Nb_3$ approach the Pauli paramagnetic limit, $\mu_0 H^{\mathrm{Pauli}} = 1.86\times T_c = 7.77$ T.

For the type-II superconductors, the destruction of superconductivity caused by the existence of a magnetic field is mainly due to two main effects: orbital limiting field and Pauli paramagnetic field. Considering their comprehensive effects, the temperature dependence of the orbital limiting field $H_{C2}^{\mathrm{orb}}(0)$ can be described by Werthamer-Helfand-Hohenberg (WHH) theory:

$$H_{C2}^{\mathrm{orb}}(0) = -\alpha T_c \left.\frac{dH_{C2}(T)}{dT}\right|_{T=T_c}$$

This equation gives the orbital limit field of the Cooper pair. The constant is 0.693 (0.73), representing the purity factor of the dirty (clean) limit superconductor. As shown in Figure S11, Supporting Information, the initial slope $-\left.\frac{dH_{C2}(T)}{dT}\right|_{T=T_c}$ yields a value of 1.66, 2.18, 1.76, 1.63, 1.65 T/K, which gives $H_{C2}^{\mathrm{orb}}(0)$ = 5.19, 6.53, 5.12, 4.74, 5.31 T, slightly smaller than the extrapolated value from the GL fitting. In order to measure the relative strength of the orbit and Pauli limit field, Maki parameters are calculated according to the formula $\alpha_M = \sqrt{2}H_{C2}^{orb}(0)/H_{C2}^{p}(0)$, which are 0.873, 1.18, 0.788, 0.755, and 0.79.

Now, various rhenium- (Re-) based NSC, such as NbReSi[41], $Re_{5.5}Ta$[42], $Re_{24}Ti_5$[43], have upper critical fields close to or even beyond the Pauli limit field, which could indicate singlet-triplet mixing. However, $Re_{3.5}Os_{3.5}Ta_{0.5}Hf_{0.5}Nb_3$, $Re_3Os_3Ta_{0.5}Hf_{0.5}Nb_3$, $Re_{3.5}Os_{3.5}Mo_{0.5}Hf_{0.5}Nb_3$, $Re_3Os_3Mo_{0.5}Hf_{0.5}Nb_3$, and $Re_{3.5}Os_{3.5}Mo_{0.5}W_{0.5}Nb_3$ cannot surpass the Pauli limit because their superconductivity is governed by conventional spin-singlet pairing and lacks the unconventional mechanisms (spin-triplet states, strong spin-orbit coupling, or topological effects) required to overcome Zeeman splitting. Despite their structural complexity and high configurational entropy, these alloys remain fundamentally constrained by the same universal limit as traditional superconductors.

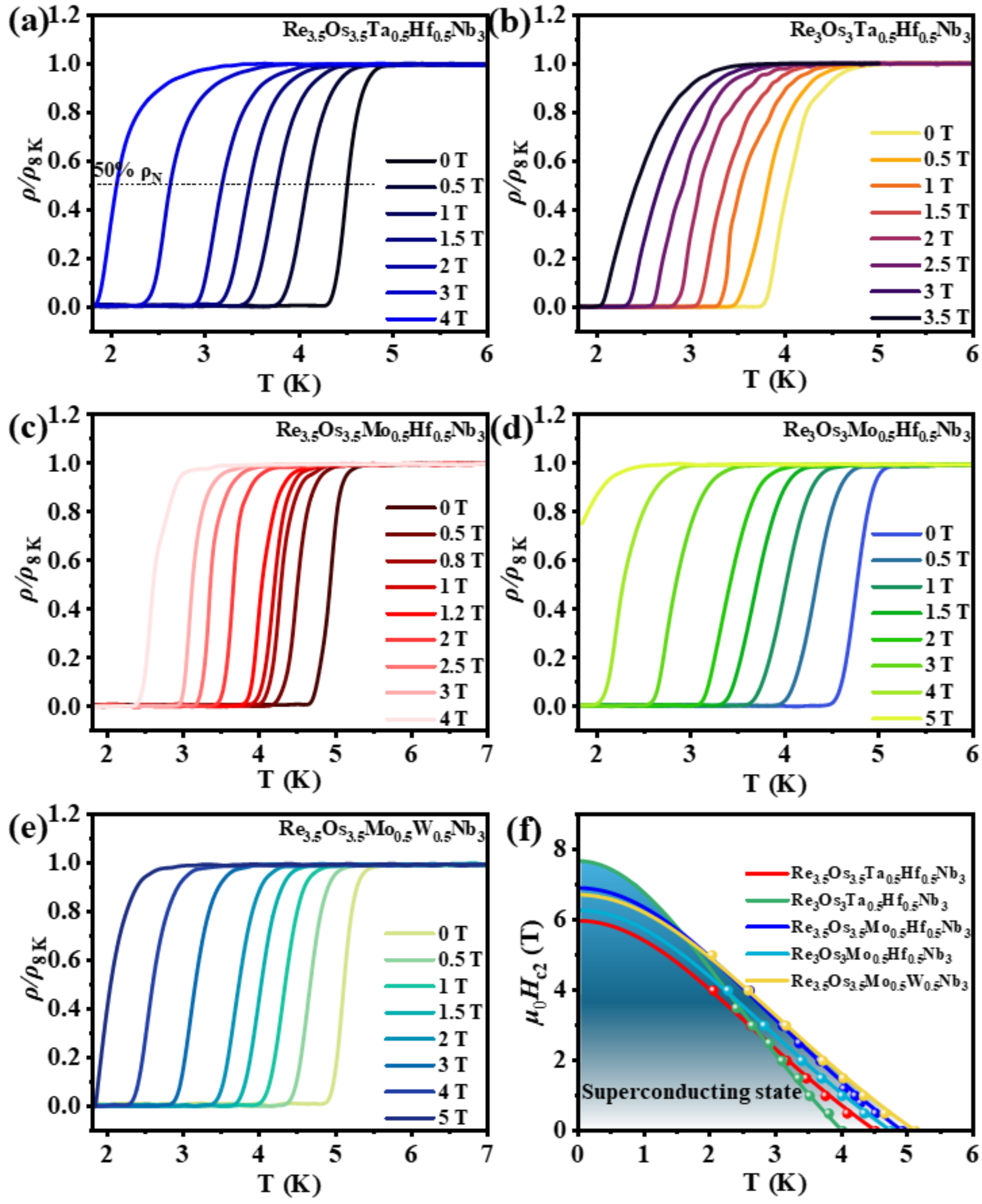

**Figure 5.** The temperature-dependent resistivity under magnetic field for (a) $Re_{3.5}Os_{3.5}Ta_{0.5}Hf_{0.5}Nb_3$, (b) $Re_3Os_3Ta_{0.5}Hf_{0.5}Nb_3$, (c) $Re_{3.5}Os_{3.5}Mo_{0.5}Hf_{0.5}Nb_3$, (d)

$Re_3Os_3Ta_{0.5}Hf_{0.5}Nb_3$, and (e) $Re_3Os_3Mo_{0.5}Hf_{0.5}Nb_3$ in different fields. (f) The upper critical field versus temperature data and the GL equation.

The GL coherence length $\xi_{GL}(0)$ can be estimated using the relation $\xi_{GL}^2(0)=\frac{\phi_0}{2\pi H_{c2}(0)}$, where $\Phi_0$ is the magnetic flux quantum. Using the values of $\mu_0H_{c2}(0)$, the estimated coherence length $\xi_{GL}(0)$ is 74 Å, 65 Å, 68 Å, 72 Å, and 70 Å for $Re_{3.5}Os_{3.5}Ta_{0.5}Hf_{0.5}Nb_3$, $Re_3Os_3Ta_{0.5}Hf_{0.5}Nb_3$, $Re_{3.5}Os_{3.5}Mo_{0.5}Hf_{0.5}Nb_3$, $Re_3Os_3Mo_{0.5}Hf_{0.5}Nb_3$, and $Re_{3.5}Os_{3.5}Mo_{0.5}W_{0.5}Nb_3$, respectively. According to the expression $\mu_0H_{c1}=\frac{\phi_0}{4\pi\lambda_{GL}^2(0)}\ln\frac{\lambda_{GL}(0)}{\xi_{GL}(0)}$, the magnetic penetration depth $\lambda_{GL}(0)$ can be calculated by using the $\mu_0H_{c1}(0)$ and $\xi_{GL}(0)$. The calculated $\lambda_{GL}(0)$ is 1133 Å, 906 Å, 896 Å, 742 Å, 593 Å for $Re_{3.5}Os_{3.5}Ta_{0.5}Hf_{0.5}Nb_3$, $Re_3Os_3Ta_{0.5}Hf_{0.5}Nb_3$, $Re_{3.5}Os_{3.5}Mo_{0.5}Hf_{0.5}Nb_3$, $Re_3Os_3Mo_{0.5}Hf_{0.5}Nb_3$, and $Re_{3.5}Os_{3.5}Mo_{0.5}W_{0.5}Nb_3$, respectively. Then, the GL parameter can be obtained as $k_{GL}=\frac{\lambda_{GL}(0)}{\xi_{GL}(0)}$ = 15.31, 13.93, 10.91, 10.30 and 8.47 > $\frac{1}{\sqrt{2}}$, affirming that $Re_{3.5}Os_{3.5}Ta_{0.5}Hf_{0.5}Nb_3$, $Re_3Os_3Ta_{0.5}Hf_{0.5}Nb_3$, $Re_{3.5}Os_{3.5}Mo_{0.5}Hf_{0.5}Nb_3$, $Re_3Os_3Mo_{0.5}Hf_{0.5}Nb_3$, and $Re_{3.5}Os_{3.5}Mo_{0.5}W_{0.5}Nb_3$ are type-II superconductors.

### 3.4 Specific Heat

To further investigate the bulk superconductivity of the alloys, specific heat measurements were performed, as shown in **Figure 6**. The $T_c$ values, determined by the apparent sharp $C_p$ jump, are 3.44 K, 2.88 K, 3.88 K, 3.39 K, 4.06 K for $Re_{3.5}Os_{3.5}Ta_{0.5}Hf_{0.5}Nb_3$, $Re_3Os_3Ta_{0.5}Hf_{0.5}Nb_3$, $Re_{3.5}Os_{3.5}Mo_{0.5}Hf_{0.5}Nb_3$, $Re_3Os_3Mo_{0.5}Hf_{0.5}Nb_3$, and $Re_{3.5}Os_{3.5}Mo_{0.5}W_{0.5}Nb_3$, respectively. The apparent sharp $C_p$ jump at $T_c$ for five samples confirms the bulk nature of superconductivity. Meanwhile, the normal-state specific heat, composed of electronic ($\gamma$) and phonon ($\beta$) contributions, can be well-fitted using the Debye model[44],

$$C_p/T=\gamma+\beta T^2+\delta T^4$$

It yields $\gamma$ = 3.58999 mJ mol$^{-1}$ K$^{-2}$, $\beta$ = 0.04847 mJ mol$^{-1}$ K$^{-4}$ for $Re_{3.5}Os_{3.5}Ta_{0.5}Hf_{0.5}Nb_3$, $\gamma$ = 3.41478 mJ mol$^{-1}$ K$^{-2}$, $\beta$ = 0.0587 mJ mol$^{-1}$ K$^{-4}$ for $Re_3Os_3Ta_{0.5}Hf_{0.5}Nb_3$, $\gamma$ = 3.31423 mJ mol$^{-1}$ K$^{-2}$, $\beta$ = 0.04208 mJ mol$^{-1}$ K$^{-4}$ for $Re_{3.5}Os_{3.5}Mo_{0.5}Hf_{0.5}Nb_3$, $\gamma$ = 3.39139 mJ mol$^{-1}$ K$^{-2}$, $\beta$ = 0.05062 mJ mol$^{-1}$ K$^{-4}$ for $Re_3Os_3Mo_{0.5}Hf_{0.5}Nb_3$, $\gamma$ = 3.79408 mJ mol$^{-1}$ K$^{-2}$, $\beta$ = 0.07692 mJ mol$^{-1}$ K$^{-4}$ for $Re_{3.5}Os_{3.5}Mo_{0.5}W_{0.5}Nb_3$. The fitting curves are shown in **Figures 6a, c, e, g, i** as solid black lines. Subsequently, the Debye temperature $\Theta_D$ can be calculated using the equation:

$$\Theta_D = \left(12\pi^4 nR/5\beta\right)^{1/3}$$

Where $n$ represents the number of atoms per formula unit. Thus, we obtain $\Theta_D$ = 342 K, 321 K, 358 K, 337 K, 293 K for $Re_{3.5}Os_{3.5}Ta_{0.5}Hf_{0.5}Nb_3$, $Re_3Os_3Ta_{0.5}Hf_{0.5}Nb_3$, $Re_{3.5}Os_{3.5}Mo_{0.5}Hf_{0.5}Nb_3$, $Re_3Os_3Mo_{0.5}Hf_{0.5}Nb_3$, and $Re_{3.5}Os_{3.5}Mo_{0.5}W_{0.5}Nb_3$, respectively.

From $\Theta_D$, we can estimate the electron-phonon coupling strength $\lambda_{ep}$ using the inverted McMillan formula,

$$\lambda_{ep} = \frac{1.04 + \mu \ln(\Theta_D / 1.45T_C)}{(1 - 0.62\mu^*)\ln(\Theta_D / 1.45T_C) - 1.04}$$

Where $\mu^*$ is the Coulomb repulsion pseudopotential, which is set to 0.13 in this paper. We obtain 0.598, 0.596, 0.605, 0.609, 0.648 for $Re_{3.5}Os_{3.5}Ta_{0.5}Hf_{0.5}Nb_3$, $Re_3Os_3Ta_{0.5}Hf_{0.5}Nb_3$, $Re_{3.5}Os_{3.5}Mo_{0.5}Hf_{0.5}Nb_3$, $Re_3Os_3Mo_{0.5}Hf_{0.5}Nb_3$, and $Re_{3.5}Os_{3.5}Mo_{0.5}W_{0.5}Nb_3$, respectively. In addition, with the formula,

$$N(E_F) = \frac{3\gamma_n}{\pi^2 k_B^2 (1 + \lambda_{ep})}$$

The electronic states at the Fermi energy $N(E_F)$ are found to be 0.95422, 0.90853, 0.87716, 0.89509, 0.97809 eV$^{-1}$ f.u.$^{-1}$ for $Re_{3.5}Os_{3.5}Ta_{0.5}Hf_{0.5}Nb_3$, $Re_3Os_3Ta_{0.5}Hf_{0.5}Nb_3$, $Re_{3.5}Os_{3.5}Mo_{0.5}Hf_{0.5}Nb_3$, $Re_3Os_3Mo_{0.5}Hf_{0.5}Nb_3$, and $Re_{3.5}Os_{3.5}Mo_{0.5}W_{0.5}Nb_3$, respectively.

According to the equation $C_{el} = C_p - \beta T^2$, after subtracting the phonon contribution, the electronic specific heat ($C_{el}$) is isolated and plotted in **Figure 6b, d, f, g, j**. The normalized specific heat jump value $\Delta C/\gamma T_c$ is found to be 1.396, 1.38, 1.484, 1.418, 1.472, which is close to the Bardeen-Cooper-Schrieffer (BCS) weak-coupling limit value of 1.43, evidencing the moderate electron-phonon coupling in these alloys.

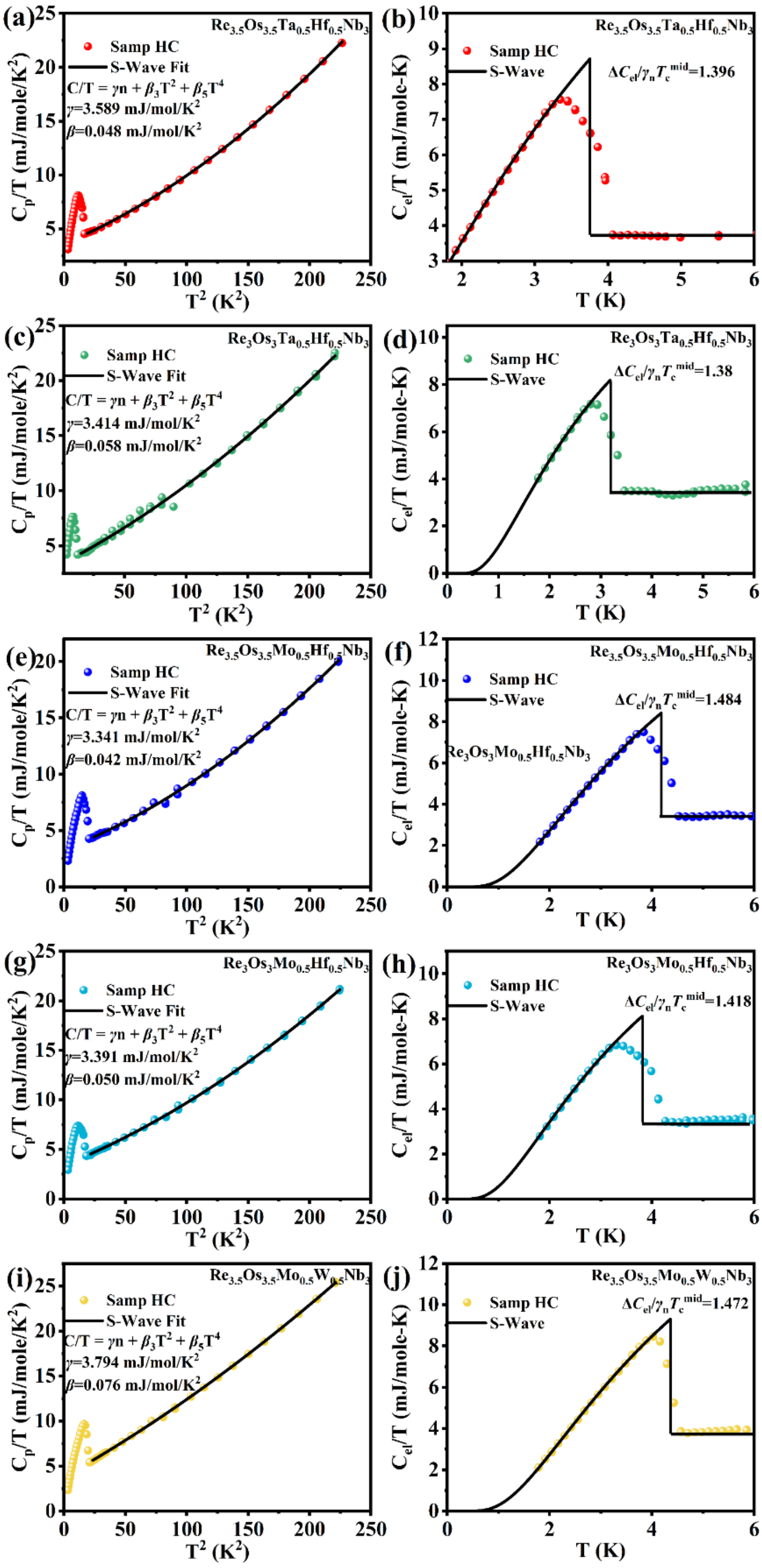

**Figure 6** (a,c, e, g, i) The $C_p$ versus $T^2$ curves, fitted with the low-temperature Debye model $C_p = \gamma + \beta T^2 + \delta T^4$. (b, d, f, h, j) $C_{el}$ versus $T$ data, the black lines are fitted with the $\alpha$-model.

**Figure 7a** shows the VEC dependence of $T_c$ for these five MEA-HEA $Re_{3.5}Os_{3.5}Ta_{0.5}Hf_{0.5}Nb_3$, $Re_3Os_3Ta_{0.5}Hf_{0.5}Nb_3$, $Re_{3.5}Os_{3.5}Mo_{0.5}Hf_{0.5}Nb_3$, $Re_3Os_3Mo_{0.5}Hf_{0.5}Nb_3$, and $Re_{3.5}Os_{3.5}Mo_{0.5}W_{0.5}Nb_3$, with related α-Mn type MEA-HEA superconductors for comparison. The blue solid line represents the variation of the $T_c$ with VEC for "classic crystalline alloy". It shows that $T_c$ peaks between VEC values of approximately 6.5 to 7.0, indicating the strongest $T_c$ of the alloy within this VEC range. The dotted lines with data points represent the experimental data for different alloy systems. These data points show the experimental results of $T_c$ for different amorphous alloys at specific VEC values. Most data points are located near or on the curve of the classic crystalline alloy, indicating a strong correlation between the superconducting performance of these alloys and VEC. As can be seen, the VEC values of HEA superconductors are between 6.45 and 6.81, suggesting that they still obey the Matthias rule. Specifically, the VEC dependency of the $T_c$s in the MEA-HEAs shows a monotonic and almost linear trend, with VEC increasing from 6.45 to 6.68 and $T_c$ also increasing accordingly, which is consistent with the trend reported previously[31]. Superconducting state parameters for our sample and other $\alpha$-Mn structure HEA systems are summarized in **Table 2**. In the $\alpha$-Mn structure MEAs-HEAs, most exhibit unconventional superconducting properties: the upper critical field exceeds the Pauli limit, moderately coupled superconductivity, and only $Re_3Os_3Ta_{0.5}Hf_{0.5}Nb_3$ closely approaches the Pauli limit. So, further comprehensive investigations on new α-Mn structure MEA-HEAs are necessary to establish a conclusive link between crystal structure, elemental composition, and superconducting pairing mechanism.

**Table S2** compares the superconducting parameters for the Re-Os MEA-HEAs superconductors with $(ZrNb)_{0.10}[MoReRu]_{0.90}$, $(HfTaWIr)_{0.40}[Re]_{0.60}$, and $(HfTaWPt)_{0.40}[Re]_{0.60}$. The $\gamma$ values of the new Re-Os MEAs-HEAs (3.31-3.79 mJ mol$^{-1}$ K$^{-2}$) are comparable to those of $(ZrNb)_{0.10}[MoReRu]_{0.90}$ (3.80 mJ mol$^{-1}$ K$^{-2}$), $(HfTaWIr)_{0.40}[Re]_{0.60}$ (3.10 mJ mol$^{-1}$ K$^{-2}$), and $(HfTaWPt)_{0.40}[Re]_{0.60}$ (2.85 mJ mol$^{-1}$ K$^{-2}$), indicating a consistent level of electronic contribution to the superconducting properties. The Debye temperatures of the new Re-Os MEAs-HEAs are higher than those of $(ZrNb)_{0.10}[MoReRu]_{0.90}$, $(HfTaWIr)_{0.40}[Re]_{0.60}$, $(HfTaWPt)_{0.40}[Re]_{0.60}$. This suggests that the Re-Os MEAs-HEAs might have a more robust phonon-mediated superconducting mechanism compared to $(ZrNb)_{0.10}[MoReRu]_{0.90}$, $(HfTaWIr)_{0.40}[Re]_{0.60}$, $(HfTaWPt)_{0.40}[Re]_{0.60}$. The upper critical field of $Re_3Os_3Ta_{0.5}Hf_{0.5}Nb_3$ and $(ZrNb)_{0.10}[MoReRu]_{0.90}$ is close to the Pauli limiting field, which could indicate singlet-triplet mixing.

To investigate the relationship between the electron-electron scattering rate and the renormalization of the electron mass, we calculated the Kadowaki-Woods ratios (KWR)[45], which assess the strength of electron-electron correlations. In the transition metals, the value of the KWR is observed to be $a_{TM}$ = 0.4 $\mu\Omega$ cm mol$^2$ K$^2$ J$^{-2}$, and $a_{HF}$ = 10 $\mu\Omega$ cm mol$^2$ K$^2$ J$^{-2}$ is the value seen in the heavy fermions. Under the premise of given $\rho_0$, $A$, $\gamma$, the KWR values of five MEA-HEA superconductors are $1.52 \times 10^{-4}$, $1.67 \times 10^{-4}$, $6.89 \times 10^{-5}$, $1.10 \times 10^{-4}$, $1.63 \times 10^{-4}$ μΩ cm mol$^2$ K$^2$ J$^{-2}$. We plotted the KWR (A versus γ) for various transition metals, heavy fermions, and other α-Mn structure MEA-HEA superconductors, as shown in **Figure 7b**. It is found that the α-Mn-type MEA-HEA superconductors have an anomalously large KWR, which is similar to other strongly correlated metals, such as transition-metal oxides, and the organic charge transfer salts. Moreover, the KWR of the α-Mn-type MEA-HEA superconductors is also larger than the transition metals and oxides, suggesting their strong electronic correlations. The large KWRs observed in the α-Mn structure MEA-HEA superconductors may originate from impurity scattering, proximity to a quantum critical point, and the proposed role of electron-phonon scattering in reduced dimensions.

The VEC values of these HEAs are plotted as functions of $T_c$, $\gamma$, and $\lambda_{ep}$ in **Figure 7c**. For HEAs, VEC tends to increase with increasing $T_c$, $\gamma$, and $\lambda_{ep}$. Note that the increase of VEC is caused by the increase of Re and Os or the replacement of Mo and W. Especially, the $T_c$ of all HEAs is almost linearly related to $\lambda_{ep}$, while the relationship with $\gamma$ is mainly controlled by electron-phonon interactions, and the relationship with $T_c$ is more dispersed.

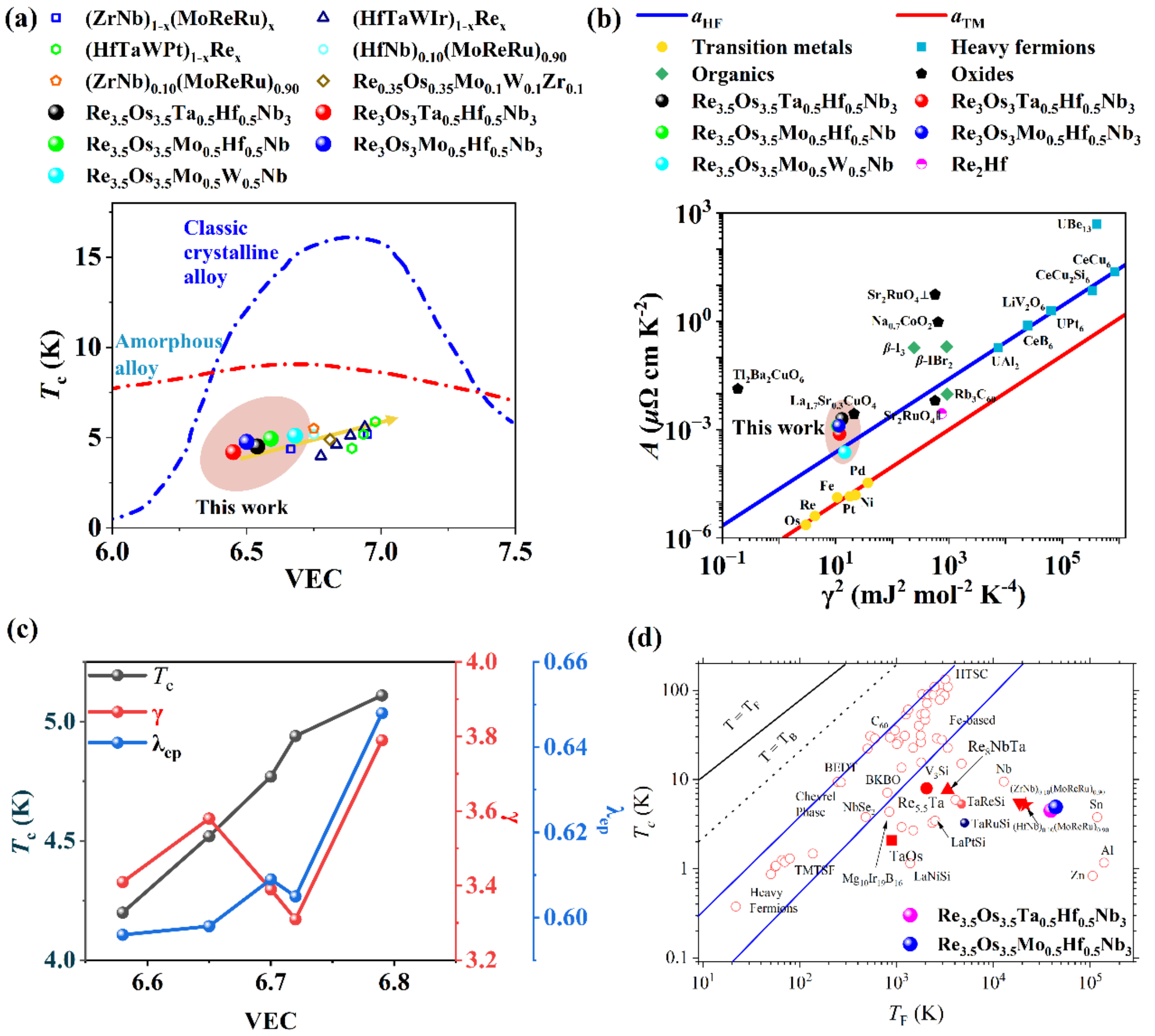

**Figure 7** (a) Theoretical VEC dependency of $T_c$ of $\alpha$-Mn type HEA-MEA superconductors. (b) The relationship between coefficient A and the Sommer-field coefficient γ for different compounds. (c) The determined VEC values of these HEAs are plotted as functions of $T_c$, $\gamma$, and $\lambda_{ep}$. (d) The Uemura plot showing the $T_c$ vs the Fermi temperature $T_F$.

For a 3D system, the Fermi temperature $T_F$ is given by the relation

$$k_B T_F = \frac{\hbar^2}{2}(3\pi^2)^{2/3}\frac{n^{2/3}}{m^*}$$

Under the premise of the given Sommerfeld coefficient $\gamma_n$ and carrier density $n$ of the quasiparticle, the effective mass of the quasiparticle $m*$ is evaluated according to the equation,

$$\gamma_n = (\frac{\pi}{3})^{2/3}\frac{k_B^2 m^* V_{f.u.} n^{1/3}}{\hbar^2 N_A}$$

By considering the calculated values of given quasiparticle number density per unit volume $n$ ($1.81 \times 10^{27}$ m$^{-3}$ for $Re_{3.5}Os_{3.5}Ta_{0.5}Hf_{0.5}Nb_3$, $1.87 \times 10^{27}$ m$^{-3}$ for $Re_{3.5}Os_{3.5}Mo_{0.5}Hf_{0.5}Nb_3$) and the effective mass of the quasiparticle $m^*$ ($0.34m_e$ for $Re_{3.5}Os_{3.5}Ta_{0.5}Hf_{0.5}Nb_3$, $0.31m_e$ for $Re_{3.5}Os_{3.5}Mo_{0.5}Hf_{0.5}Nb_3$), the estimated value of Fermi temperature $T_F$ = 39214 and 43885 K for $Re_{3.5}Os_{3.5}Ta_{0.5}Hf_{0.5}Nb_3$ and $Re_{3.5}Os_{3.5}Mo_{0.5}Hf_{0.5}Nb_3$. $T_c/T_F$ ratios are 0.000112 and 0.000115 for $Re_{3.5}Os_{3.5}Ta_{0.5}Hf_{0.5}Nb_3$ and $Re_{3.5}Os_{3.5}Mo_{0.5}Hf_{0.5}Nb_3$. According to Uemura et al., the range $0.01 \leq T_c/T_F \leq 0.1$ is defined as an unconventional band. However, $Re_{3.5}Os_{3.5}Ta_{0.5}Hf_{0.5}Nb_3$ and $Re_{3.5}Os_{3.5}Mo_{0.5}Hf_{0.5}Nb_3$ are far from the boundary of the unconventional superconductor as shown in **Figure 7d**, like the other noncentrosymmetric and unconventional superconductors such as: $Re_{5.5}Ta$, $Re_8NbTa$, $(HfNb)_{0.10}(MoReRu)_{0.90}$, and $(ZrNb)_{0.10}(MoReRu)_{0.90}$.

## 4. Conclusion

In summary, we have designed and synthesized five previously unreported Re-Os-based α-Mn-type MEAs-HEAs: $Re_{3.5}Os_{3.5}Ta_{0.5}Hf_{0.5}Nb_3$, $Re_3Os_3Ta_{0.5}Hf_{0.5}Nb_3$, $Re_{3.5}Os_{3.5}Mo_{0.5}Hf_{0.5}Nb_3$, $Re_3Os_3Mo_{0.5}Hf_{0.5}Nb_3$, and $Re_{3.5}Os_{3.5}Mo_{0.5}W_{0.5}Nb_3$. XRD results confirmed that all five MEAs-HEAs have an NC α-Mn structure. The combination of electrical transport, magnetization, and specific heat measurements indicates that all these MEAs-HEAs are superconductors with the $T_c$s ranging from 4.20 K to 5.11 K. Our analyses confirm that five MEAs-HEAs exhibit type-II superconductivity, and the superconducting properties of MEAs-HEAs remained robust even after immersion in the acid environment for over one month. The $T_c$s increase with increasing VEC in these α-Mn type HEA superconductors, which provides a clue for further design of new NC HEA superconductors. Moreover, a strong electron correlation in these α-Mn structure HEA superconductors, which to that in typical heavy fermion superconductors, is revealed by a large KWR. On the other hand, since many α-Mn-type Re-based binary superconductors exhibit time-reversal symmetry breaking, thus, it will be interesting to further study whether this unconventional feature will occur in these Re-based MHA-HEA superconductors.

## Acknowledgment

This work is supported by the National Natural Science Foundation of China (12274471, 12404165, 11922415), Guangzhou Science and Tech-nology Programme (No. 2024A04J6415), and the State Key Laboratory of Optoelectronic Materials and Technologies (Sun Yat-Sen University, No. OEMT-2024-ZRC-02), Key Laboratory of Magnetoelectric Physics and Devices of Guangdong Province (Grant No. 2022B1212010008) and Research Center for Magnetoelectric Physics of Guangdong Province (2024B0303390001). Lingyong Zeng acknowledges the Postdoctoral Fellowship Program of CPSF (GZC20233299) and the Fundamental Research Funds for the Central Universities, Sun Yat-sen University (24qupy092).

**Table 2.** The superconducting properties of the $\alpha$-Mn structure HEA superconductors

| Parameters | unit | $Re_{3.5}Os_{3.5}Ta_{0.5}Hf_{0.5}Nb_3$ | $Re_3Os_3Ta_{0.5}Hf_{0.5}Nb_3$ | $Re_{3.5}Os_{3.5}Mo_{0.5}Hf_{0.5}Nb_3$ | $Re_3Os_3Mo_{0.5}Hf_{0.5}Nb_3$ | $Re_{3.5}Os_{3.5}Mo_{0.5}W_{0.5}Nb_3$ | $Re_{0.35}Os_{0.35}Mo_{0.1}W_{0.1}Zr_{0.1}$ | $(HfNb)_{0.10}(MoReRu)_{0.90}$ | $(ZrNb)_{0.10}(MoReRu)_{0.90}$ |
|---|---|---|---|---|---|---|---|---|---|
| $T_c$ | K | 4.52 | 4.20 | 4.94 | 4.77 | 5.11 | 4.90 | 5.2 | 5.5 |
| VEC | e/a | 6.54 | 6.45 | 6.59 | 6.5 | 6.68 | 6.81 | | |
| $\mu_0 H^*_{c1}(0)$ | mT | 35 | 53 | 53 | 70 | 100 | 4.04 | 2.1 | 3.2 |
| $\mu_0 H^{GL}_{C2}(0)$ | T | 5.99 | 7.71 | 6.91 | 6.29 | 6.69 | 9.10 | 9.56 | 10.12 |
| $\mu_0 H_P(0)$ | T | 8.4072 | 7.812 | 9.1884 | 8.8722 | 9.5046 | 9.604 | 9.672 | 10.23 |
| $\mu_0 H^{orb}(0)$ | T | 5.19 | 6.53 | 5.12 | 4.74 | 5.31 | | | |
| $\mu_0 H^{GL}_{C2}(0)/\mu_0 H_P(0)$ | | 0.74 | 0.98 | 0.75 | 0.70 | 0.70 | 0.91 | 0.98 | 0.98 |
| $\xi_{GL}$ | Å | 74 | 65 | 68 | 72 | 70 | 53.0 | 59.2 | 53.5 |
| $\lambda_{GL}$ | Å | 1133 | 906 | 896 | 742 | 593 | 4280 | 6090 | 4870 |
| $k_{GL}$ | | 15.31 | 13.93 | 10.91 | 10.30 | 8.47 | 81 | 103 | 91 |
| $\gamma$ | mJ mol$^{-1}$ K$^{-2}$ | 3.58 | 3.41 | 3.31 | 3.39 | 3.79 | 2.52 | 3.6 | 3.8 |
| $\theta_D$ | K | 342 | 321 | 358 | 337 | 293 | 560 | 346 | 335 |
| $\lambda_{ep}$ | | 0.598 | 0.596 | 0.605 | 0.609 | 0.648 | 0.53 | 0.62 | 0.63 |
| $\boldsymbol{N(E_F)}$ | $eV^{-1}$ f.u.$^{-1}$ | 0.95422 | 0.90853 | 0.87716 | 0.89509 | 0.97809 | | | |
| $\boldsymbol{\rho_0}$ | $\mu\Omega$ cm | $-3.43\times10^{-5}$ | $1.67\times10^{-4}$ | $6.89\times10^{-5}$ | $1.10\times10^{-4}$ | $2.88\times10^{-5}$ | | 204 | 238 |
| $\boldsymbol{\Delta C/\gamma T_c}$ | | 1.396 | 1.38 | 1.484 | 1.418 | 1.472 | | 1.67 | 1.49 |

# Supporting Information

## Strongly correlated electronic superconductivity in the noncentrosymmetric Re-Os-based high/medium-entropy alloys

*Rui Chen*[1], *Longfu Li*[1], *Lingyong Zeng*[1,2], *Kuan Li*[1], *Peifeng Yu*[1], *Kangwang Wang*[1], *Zaichen Xiang*[1], *Shuangyue Wang*[1], *Jingjun Qin*[1], *Wanyi Zhang*[1], *Yucheng Li*[1], *Tian Shang*[3], *Huixia Luo*[1, 4,5,6*]

[1]*School of Materials Science and Engineering, Sun Yat-sen University, No. 135, Xingang Xi Road, Guangzhou, 510275, P. R. China*

[2]*Device Physics of Complex Materials, Zernike Institute for Advanced Materials, University of Groningen, Nijenborgh 4, 9747 AG Groningen, The Netherlands*

[3]*Key Laboratory of Polar Materials and Devices (MOE), School of Physics and Electronic Science, East China Normal University, Shanghai 200241, China*

[4]*State Key Laboratory of Optoelectronic Materials and Technologies, Sun Yat-sen University, No. 135, Xingang Xi Road, Guangzhou, 510275, P. R. China*

[5]*Key Lab of Polymer Composite & Functional Materials, Sun Yat-sen University, No. 135, Xingang Xi Road, Guangzhou, 510275, P. R. China*

[6]*Guangdong Provincial Key Laboratory of Magnetoelectric Physics and Devices, Sun Yat-sen University, No. 135, Xingang Xi Road, Guangzhou, 510275, P. R. China*

[*]*Corresponding author. Email: luohx7@mail.sysu.edu.cn (H. Luo)*

**Table S1** Reliability factors ($R_{wp}$, $R_p$, $\chi^2$) from the Rietveld fits for all five alloys.

| | $R_{wp}$ | $R_p$ | $\chi^2$ |
|---|---|---|---|
| $Re_{3.5}Os_{3.5}Ta_{0.5}Hf_{0.5}Nb_3$ | 2.37 | 1.69 | 1.17 |
| $Re_3Os_3Ta_{0.5}Hf_{0.5}Nb_3$ | 5.64 | 3.93 | 6.07 |
| $Re_{3.5}Os_{3.5}Mo_{0.5}Hf_{0.5}Nb_3$ | 5.53 | 3.82 | 5.59 |
| $Re_3Os_3Mo_{0.5}Hf_{0.5}Nb_3$ | 7.42 | 5.56 | 1.09 |
| $Re_{3.5}Os_{3.5}Mo_{0.5}W_{0.5}Nb_3$ | 7.01 | 5.36 | 0.919 |

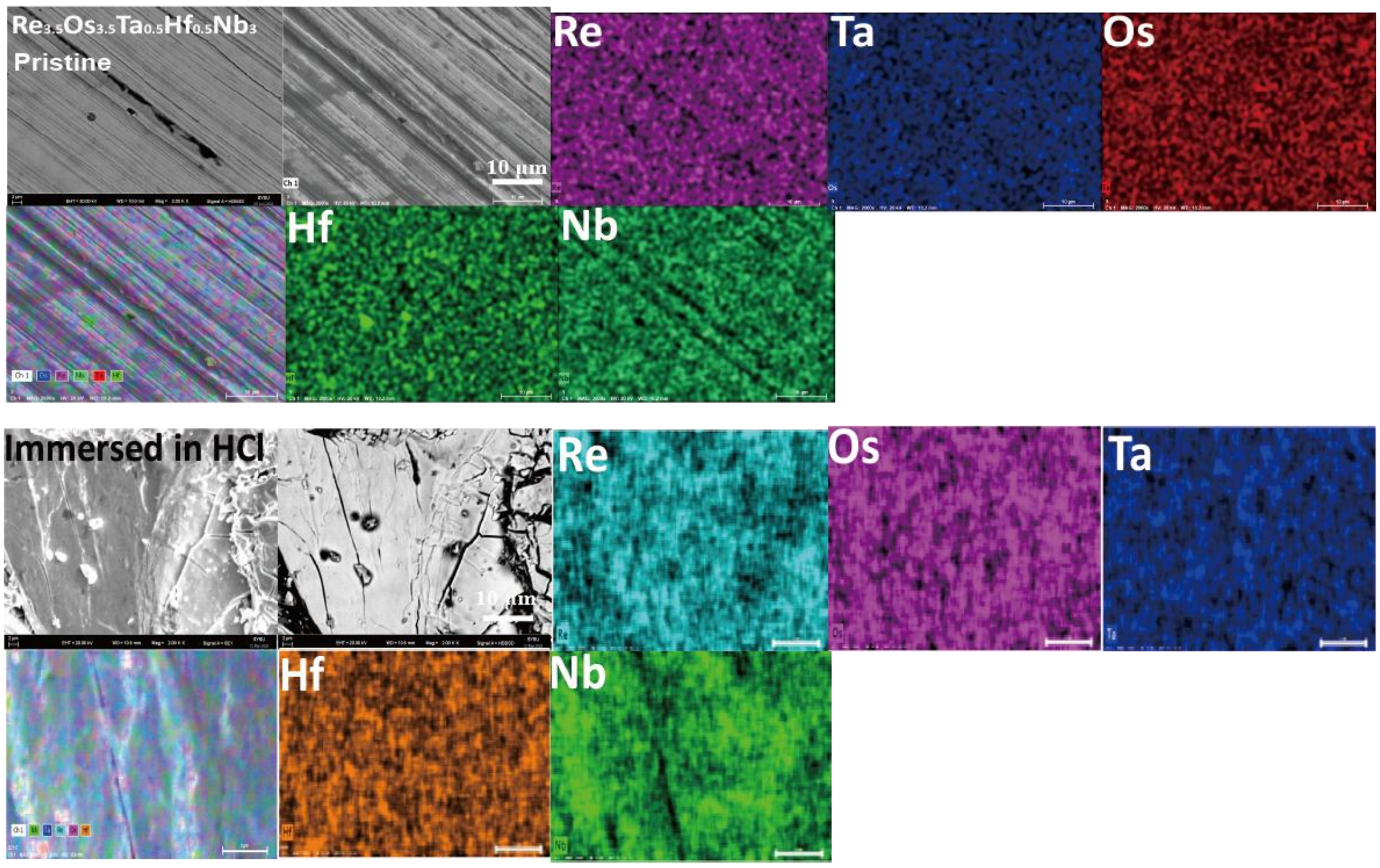

**Figure S1** SEM and BSEM images of pristine $Re_{3.5}Os_{3.5}Ta_{0.5}Hf_{0.5}Nb_3$ and spent samples after being immersed in HCl solution for one month, respectively.

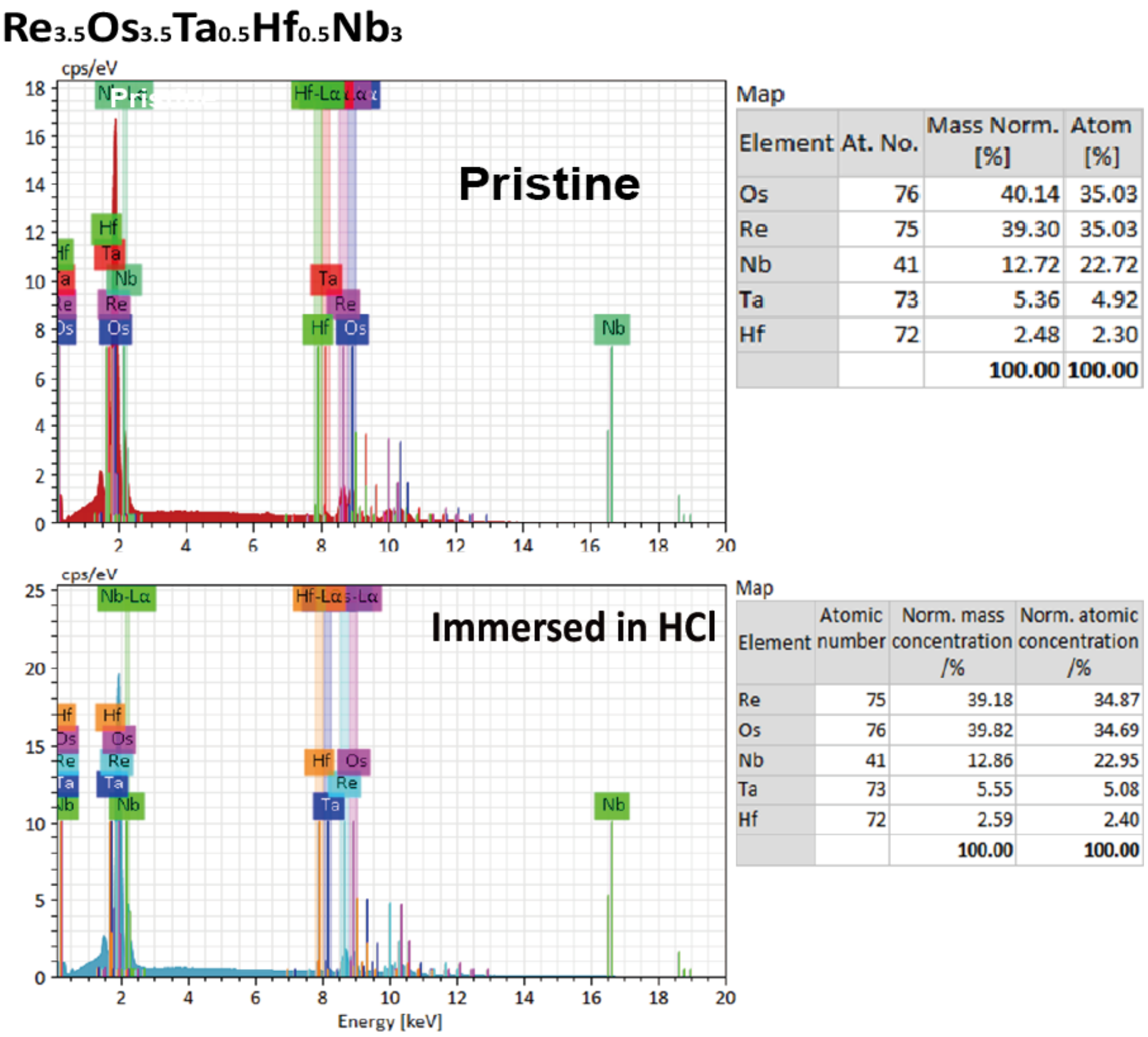

| Element | At. No. | Mass Norm. [%] | Atom [%] |
|---|---|---|---|
| Os | 76 | 40.14 | 35.03 |
| Re | 75 | 39.30 | 35.03 |
| Nb | 41 | 12.72 | 22.72 |
| Ta | 73 | 5.36 | 4.92 |
| Hf | 72 | 2.48 | 2.30 |
| | | 100.00 | 100.00 |

| Element | Atomic number | Norm. mass concentration /% | Norm. atomic concentration /% |
|---|---|---|---|
| Re | 75 | 39.18 | 34.87 |
| Os | 76 | 39.82 | 34.69 |
| Nb | 41 | 12.86 | 22.95 |
| Ta | 73 | 5.55 | 5.08 |
| Hf | 72 | 2.59 | 2.40 |
| | | 100.00 | 100.00 |

**Figure S2** EDS elemental mappings of pristine $Re_{3.5}Os_{3.5}Ta_{0.5}Hf_{0.5}Nb_3$ and spent samples after being immersed in HCl solution for one month, respectively.

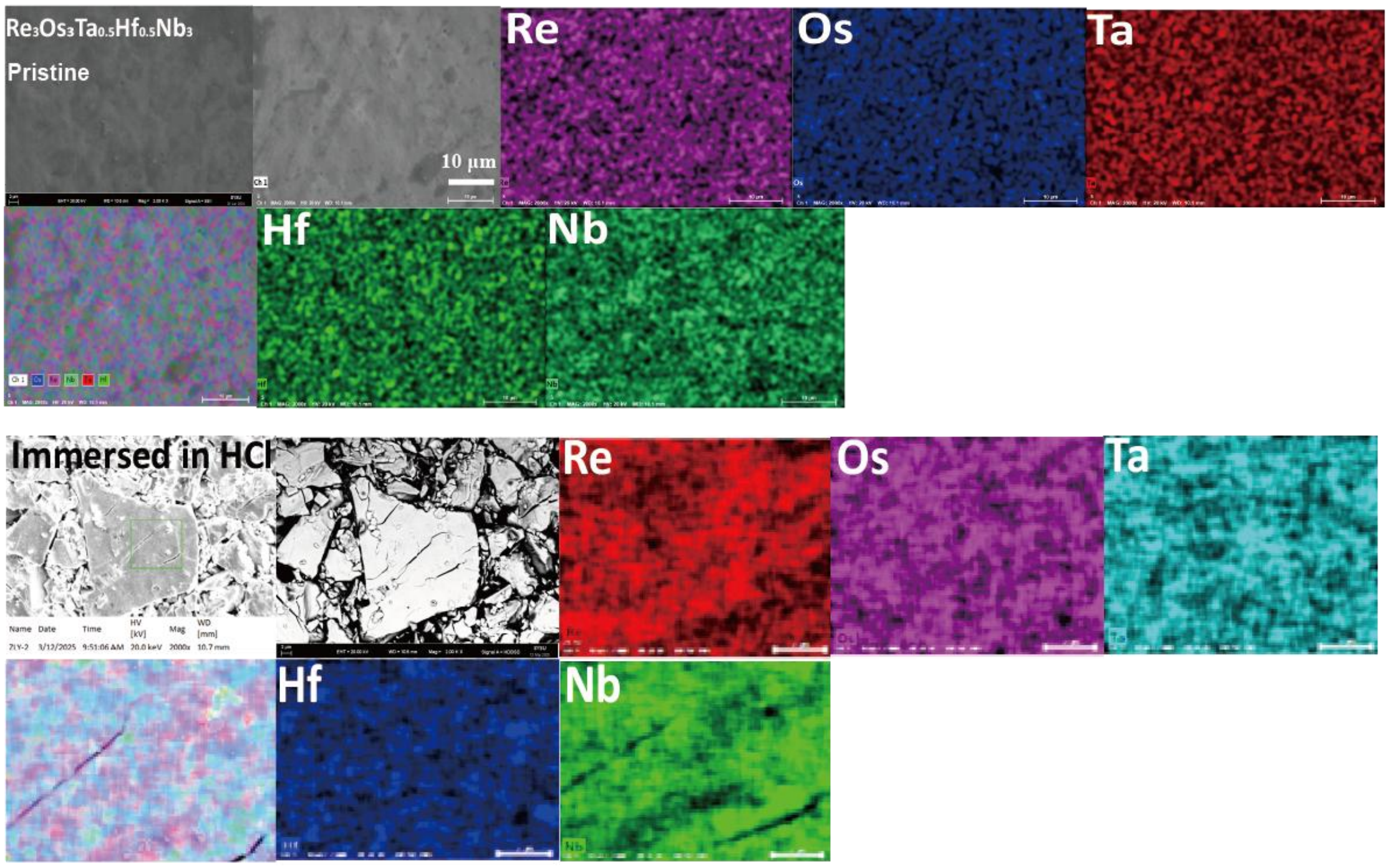

**Figure S3** SEM and BSEM images of pristine $Re_3Os_3Ta_{0.5}Hf_{0.5}Nb_3$ and spent samples after being immersed in HCl solution for one month, respectively.

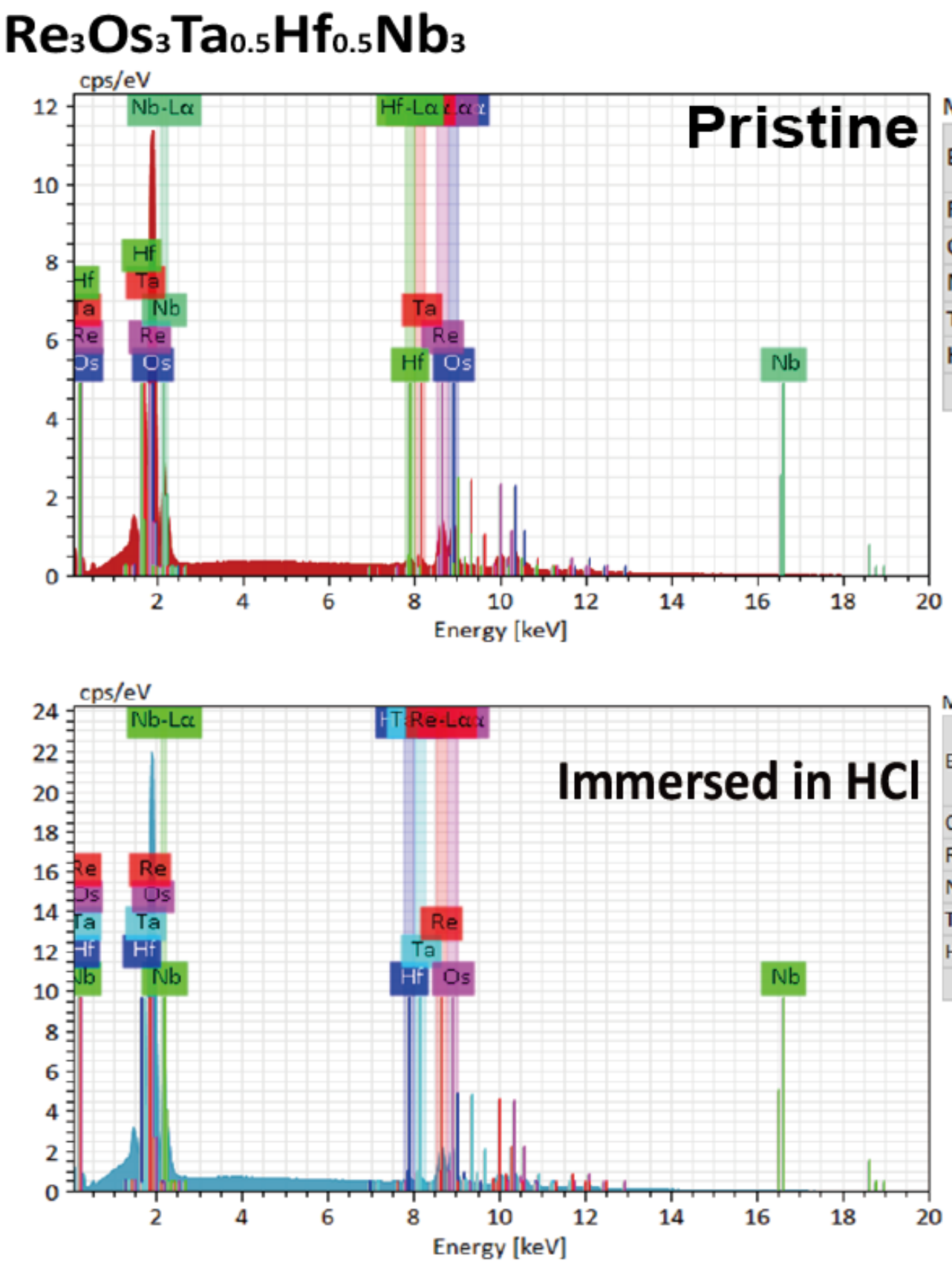

| Element | At. No. | Mass Norm. [%] | Atom [%] |
|---|---|---|---|
| Re | 75 | 37.26 | 32.50 |
| Os | 76 | 37.73 | 32.22 |
| Nb | 41 | 15.04 | 26.29 |
| Ta | 73 | 5.98 | 5.37 |
| Hf | 72 | 3.98 | 3.62 |
| | | 100.00 | 100.00 |

| Element | Atomic number | Norm. mass concentration /% | Norm. atomic concentration /% |
|---|---|---|---|
| Os | 76 | 38.10 | 32.33 |
| Re | 75 | 35.92 | 31.14 |
| Nb | 41 | 15.74 | 27.34 |
| Ta | 73 | 6.01 | 5.36 |
| Hf | 72 | 4.23 | 3.83 |
| | | 100.00 | 100.00 |

**Figure S4** EDS elemental mappings of pristine $Re_3Os_3Ta_{0.5}Hf_{0.5}Nb_3$ and spent samples after being immersed in HCl solution for one month, respectively.

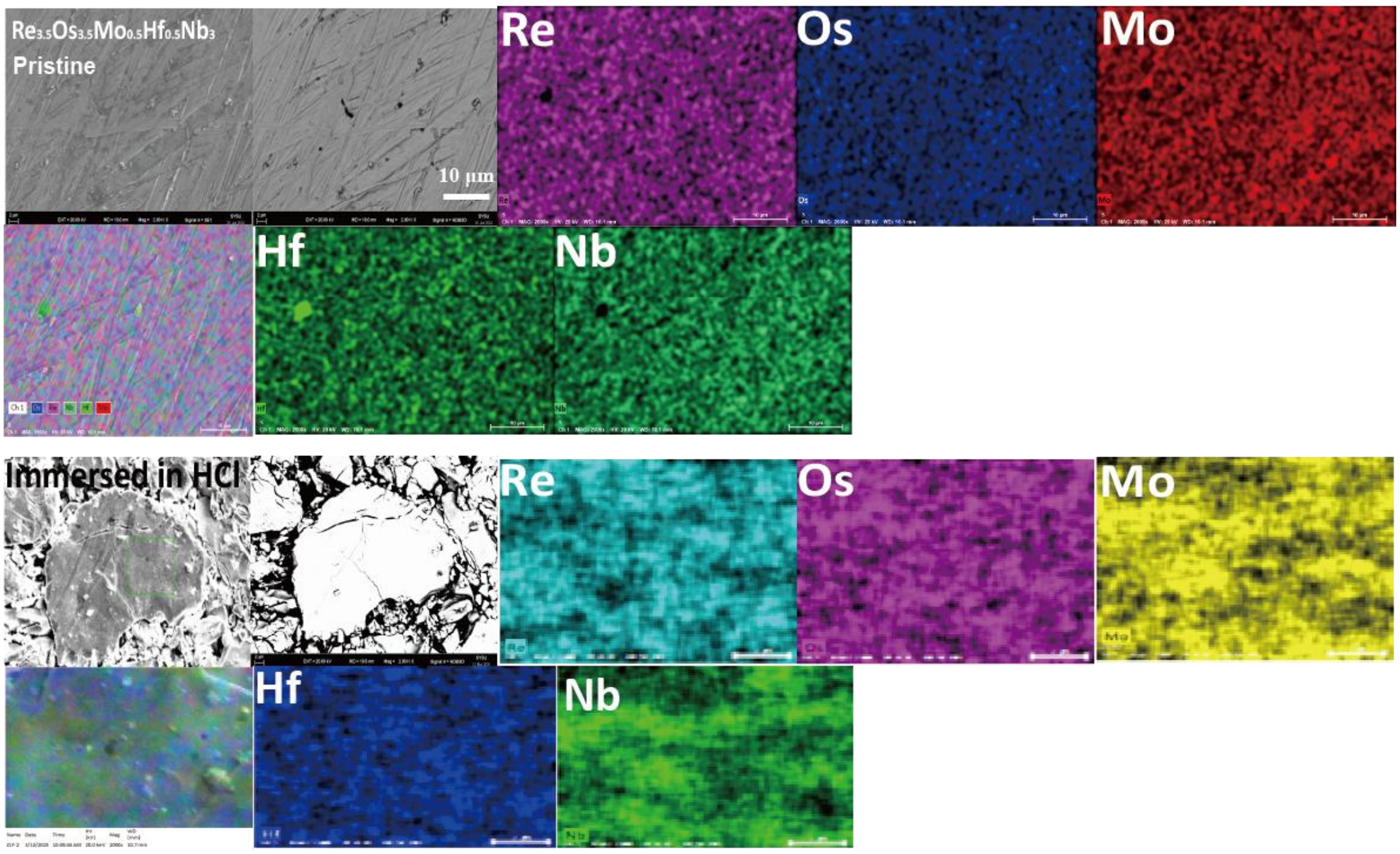

**Figure S5** SEM and BSEM images of pristine $Re_{3.5}Os_{3.5}Mo_{0.5}Hf_{0.5}Nb_3$ and spent samples after being immersed in HCl solution for one month, respectively.

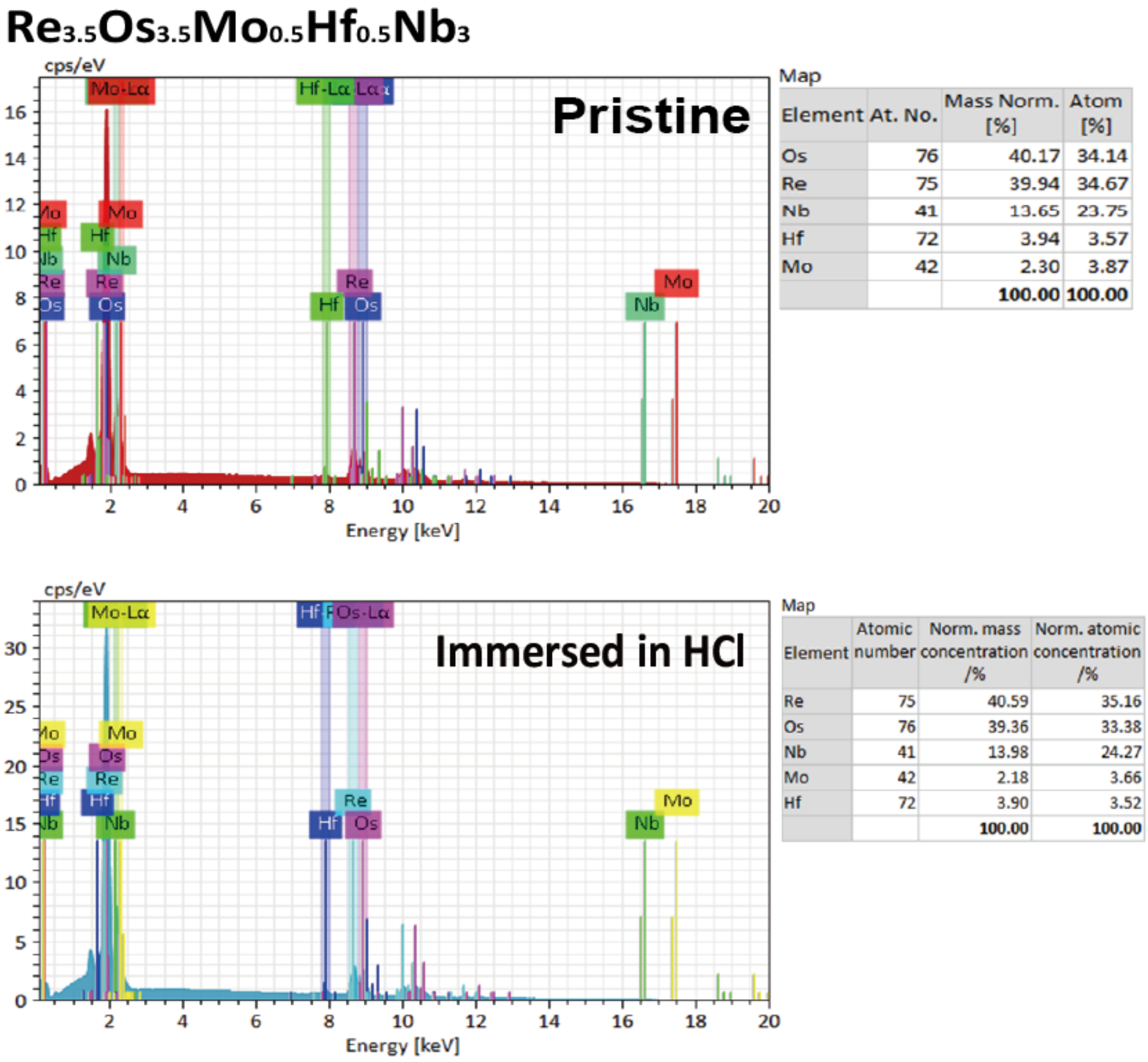

**Figure S6** EDS elemental mappings of pristine $Re_{3.5}Os_{3.5}Mo_{0.5}Hf_{0.5}Nb_3$ and spent samples after being immersed in HCl solution for one month, respectively.

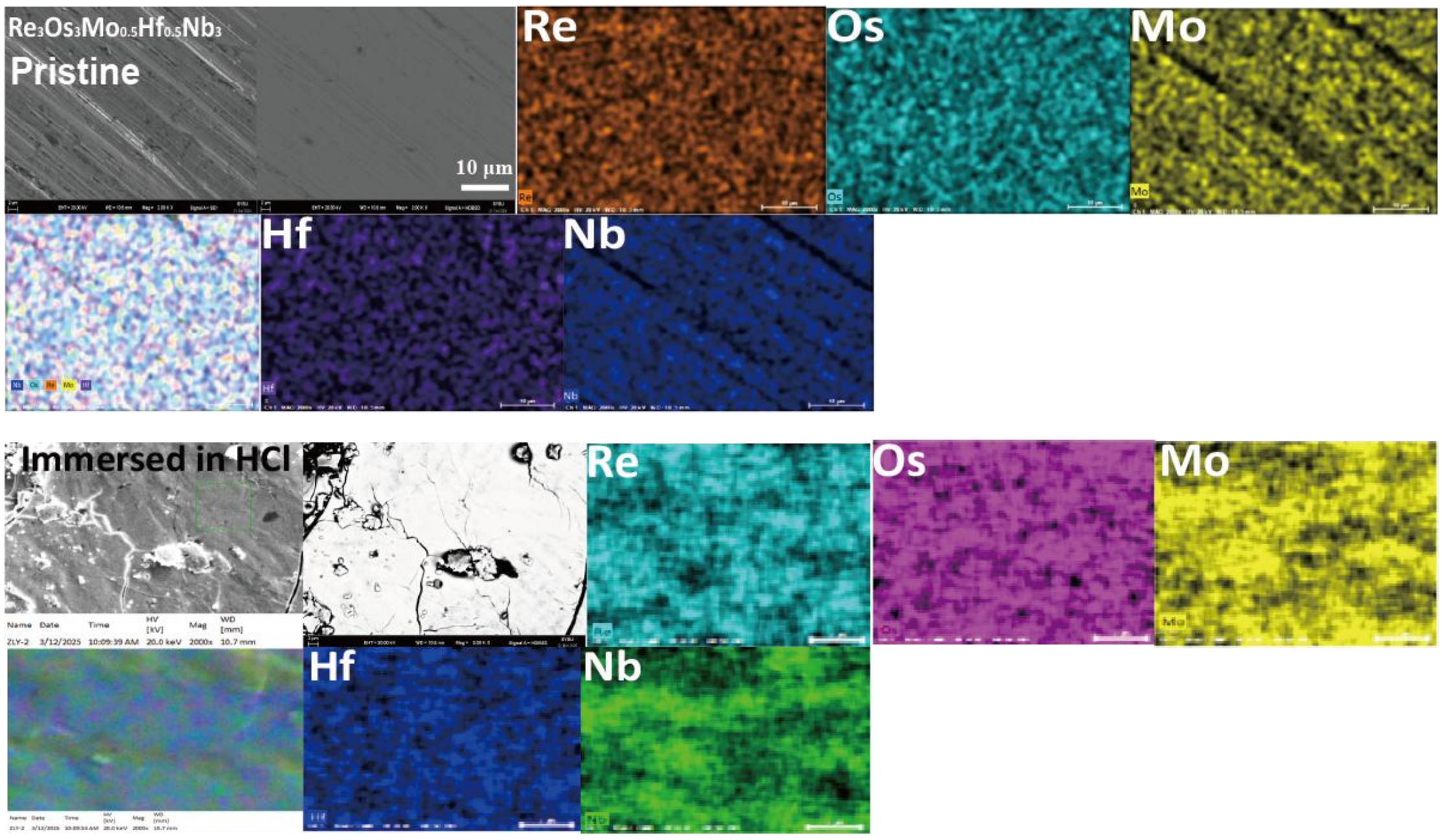

**Figure S7** SEM and BSEM images of pristine $Re_3Os_3Mo_{0.5}Hf_{0.5}Nb_3$ and spent samples after being immersed in HCl solution for one month, respectively.

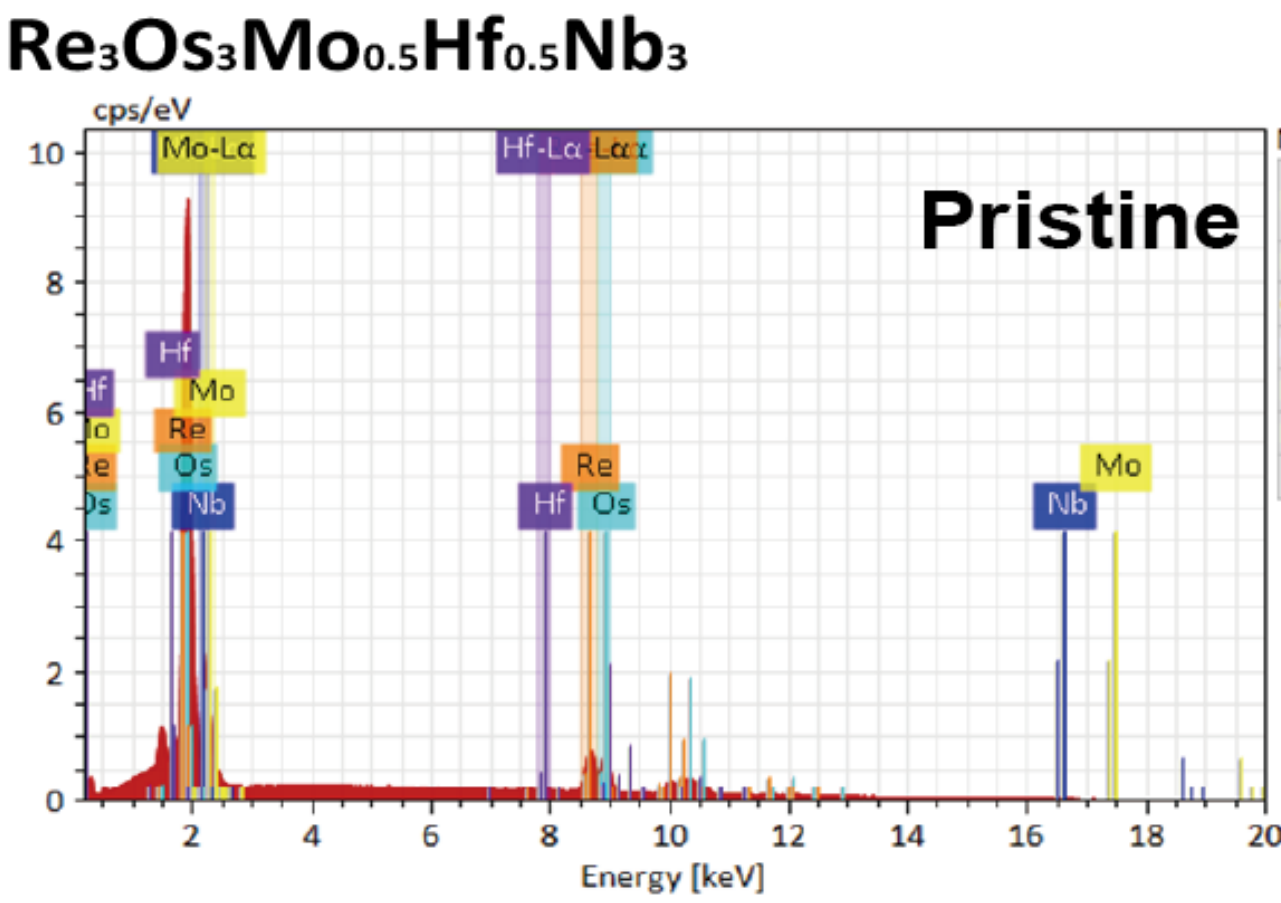

Map

| Element | At. No. | Mass Norm. [%] | Atom [%] |
|---|---|---|---|
| Re | 75 | 38.79 | 32.81 |
| Os | 76 | 39.21 | 32.46 |
| Nb | 41 | 15.90 | 26.94 |
| Mo | 42 | 3.17 | 5.20 |
| Hf | 72 | 2.94 | 2.59 |
| | | 100.00 | 100.00 |

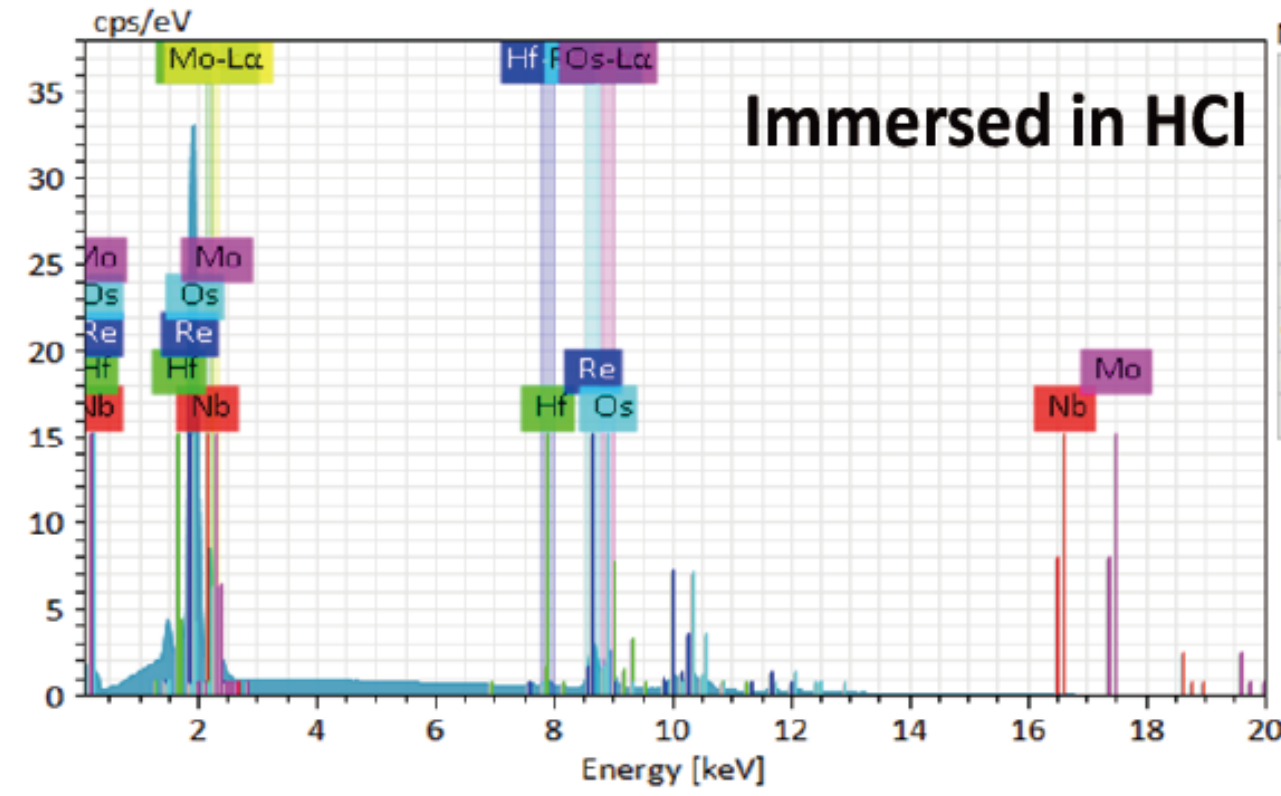

Map

| Element | Atomic number | Norm. mass concentration /% | Norm. atomic concentration /% |
|---|---|---|---|
| Nb | 41 | 15.12 | 25.87 |
| Hf | 72 | 2.79 | 2.48 |
| Re | 75 | 41.03 | 35.02 |
| Os | 76 | 38.24 | 31.96 |
| Mo | 42 | 2.81 | 4.66 |
| | | 100.00 | 100.00 |

**Figure S8** EDS elemental mappings of pristine $Re_3Os_3Mo_{0.5}Hf_{0.5}Nb_3$ and spent samples after being immersed in HCl solution for one month, respectively.

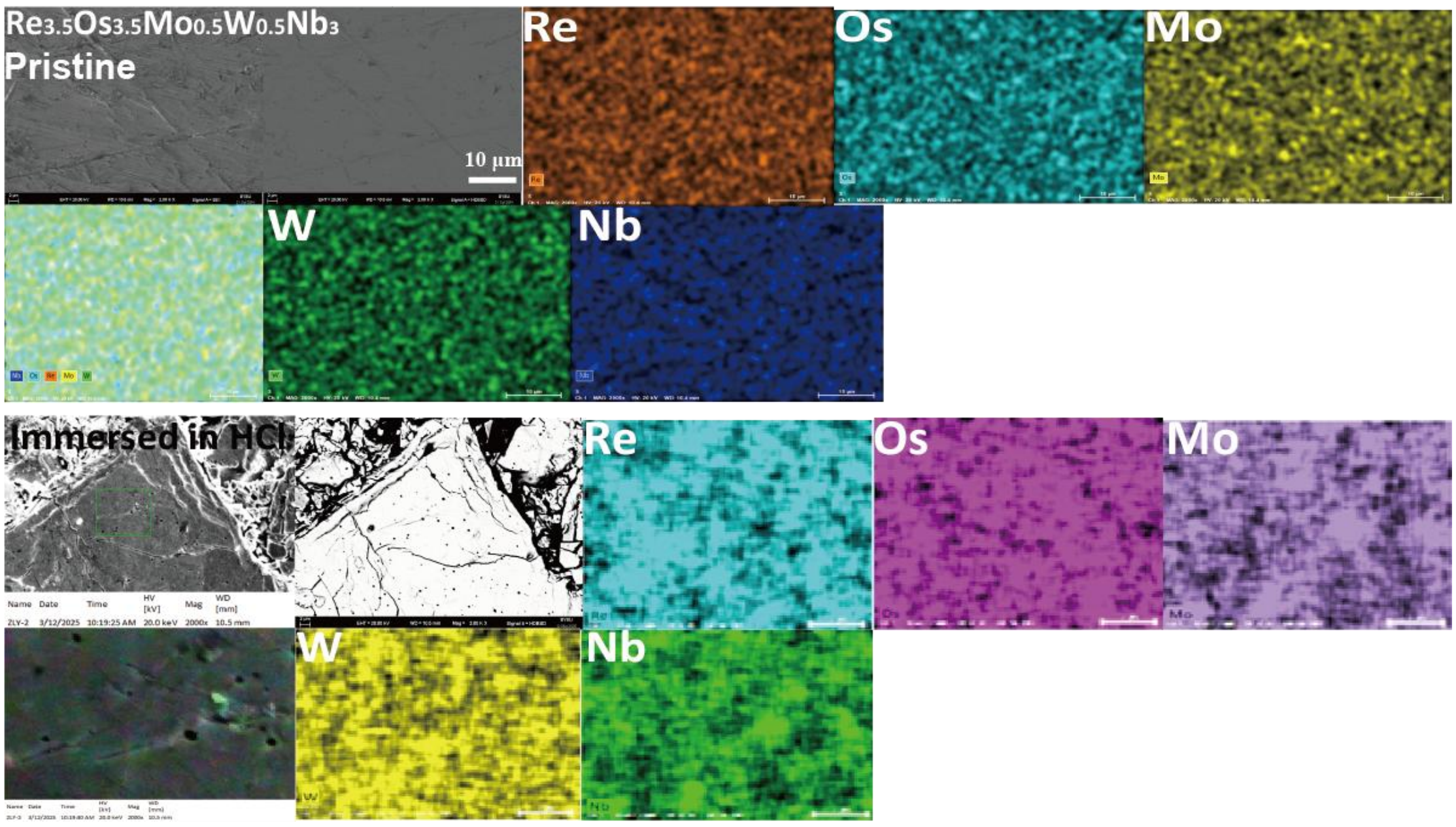

**Figure S9** SEM and BSEM images of pristine $Re_{3.5}Os_{3.5}Mo_{0.5}W_{0.5}Nb_3$ and spent samples after being immersed in HCl solution for one month, respectively.

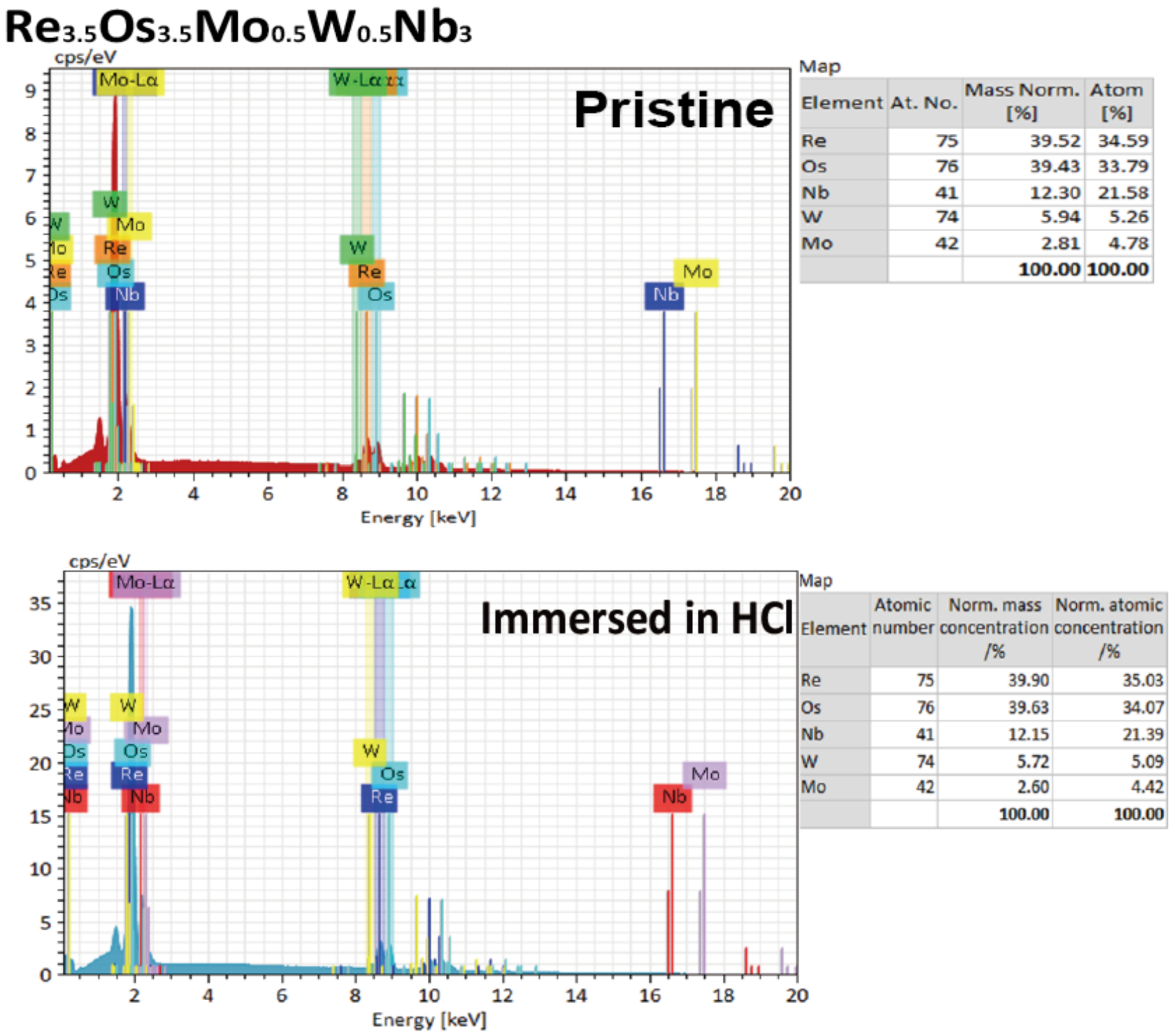

| Element | At. No. | Mass Norm. [%] | Atom [%] |
|---|---|---|---|
| Re | 75 | 39.52 | 34.59 |
| Os | 76 | 39.43 | 33.79 |
| Nb | 41 | 12.30 | 21.58 |
| W | 74 | 5.94 | 5.26 |
| Mo | 42 | 2.81 | 4.78 |
| | | 100.00 | 100.00 |

| Element | Atomic number | Norm. mass concentration /% | Norm. atomic concentration /% |
|---|---|---|---|
| Re | 75 | 39.90 | 35.03 |
| Os | 76 | 39.63 | 34.07 |
| Nb | 41 | 12.15 | 21.39 |
| W | 74 | 5.72 | 5.09 |
| Mo | 42 | 2.60 | 4.42 |
| | | 100.00 | 100.00 |

**Figure S10** EDS elemental mappings of pristine $Re_{3.5}Os_{3.5}Mo_{0.5}W_{0.5}Nb_3$ and spent samples after being immersed in HCl solution for one month, respectively.

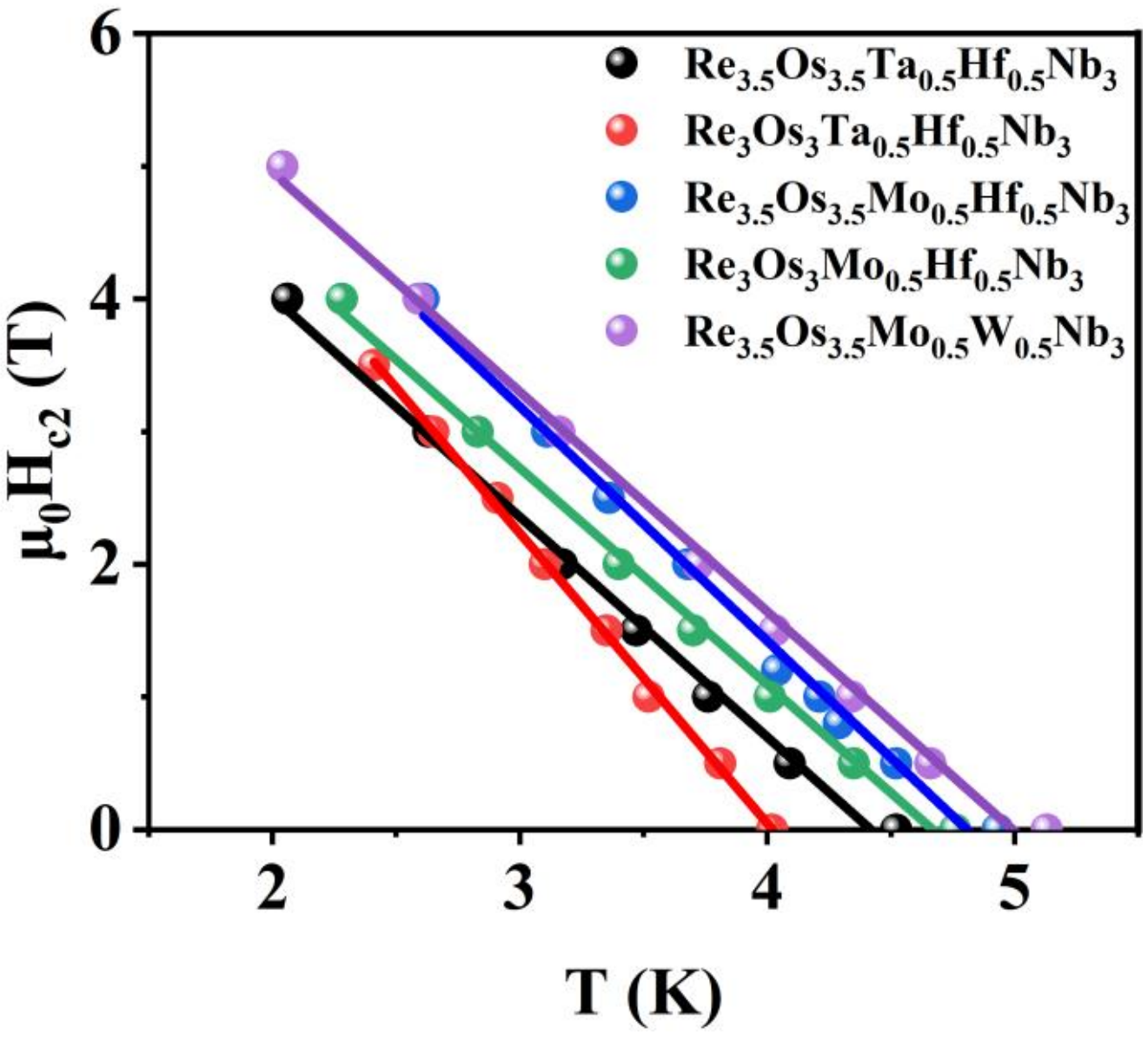

**Figure S11** The temperature dependence of the upper critical field $\mu_0H_{c2}$ with the WHH model fitting.

**Table S2** The superconducting parameters for the α-Mn-type HEAs-MEAs superconductors.

| Parameter | $Re_{3.5}Os_{3.5}Ta_{0.5}Hf_{0.5}Nb_3$ | $Re_3Os_3Ta_{0.5}Hf_{0.5}Nb_3$ | $Re_{3.5}Os_{3.5}Mo_{0.5}Hf_{0.5}Nb_3$ | $Re_3Os_3Mo_{0.5}Hf_{0.5}Nb_3$ | $Re_{3.5}Os_{3.5}Mo_{0.5}W_{0.5}Nb_3$ | $(ZrNb)_{0.10}[MoReRu]_{0.90}$ | $(HfTaWIr)_{0.40}[Re]_{0.60}$ | $(HfTaWPt)_{0.40}[Re]_{0.60}$ |
|---|---|---|---|---|---|---|---|---|
| $T_c$ (K) | 4.52 | 4.20 | 4.94 | 4.77 | 5.11 | 5.3 | 4.0 | 4.4 |
| $H_{c2}$ | 5.99 | 7.71 | 6.91 | 6.29 | 6.69 | 7.86 | 4.64 | 5.90 |
| $\gamma$ (mJ·mol$^{-1}$·K$^{-2}$) | 3.58 | 3.41 | 3.31 | 3.39 | 3.79 | 3.80(1) | 3.10(1) | 2.85(1) |
| $\beta$ (mJ·mol$^{-1}$·K$^{-4}$) | 0.04847 | 0.0587 | 0.04208 | 0.05062 | 0.07692 | 0.050(1) | 0.061(1) | 0.121(1) |
| $\Theta_D$ (K) | 342 | 321 | 358 | 337 | 293 | 339 | 317 | 252 |
| $\Delta C/\gamma \cdot T_c$ | 1.396 | 1.38 | 1.484 | 1.418 | 1.472 | 1.53 | 1.46 | 1.46 |